\documentclass[12pt]{article}
\usepackage[cp1251]{inputenc}
\usepackage[russian]{babel}
\usepackage{graphicx}
\usepackage{amsmath,amssymb}
\usepackage{caption}

\def\la{\mathrel{\mathpalette\fun <}}
\def\ga{\mathrel{\mathpalette\fun >}}
\def\fun#1#2{\lower3.6pt\vbox{\baselineskip0pt\lineskip.9pt
\ialign{$\mathsurround=0pt#1\hfil##\hfil$\crcr#2\crcr\sim\crcr}}}

\title{{\bf Современное состояние физики элементарных частиц} \\
{\normalsize Лекции на Высшей школе физики Госкорпорации Росатом} \\
{\small Саров, 15-21 апреля 2013 года}}
\author{М.И. Высоцкий \\ {\small ИТЭФ им. А.И. Алиханова, Москва}}
\date{}

\begin{document}

\maketitle

\begin{center}

{\bf План}

\end{center}

Лекция 1. Введение: фундаментальные частицы, законы сохранения,
ускорители, 4 взаимодействия, стандартная $SU(3)_c \times SU(2)_L
\times U(1)$ модель.

Лекция 2. $U(1)$ -- квантовая электродинамика (QED).

Лекция 3. $SU(3)_c$ -- квантовая хромодинамика (QCD).

Лекция 4. $SU(2)_L \times U(1)$ -- электрослабая теория.

Лекция 5. Нейтрино.

Лекция 6. За Стандартной Моделью.

\newpage

\begin{center}

{\bf Лекция 1 \\
Введение: фундаментальные частицы, законы сохранения, ускорители,
4 взаимодействия, стандартная $SU(3)_c \times SU(2)_L \times U(1)$
модель.}

\end{center}

\setcounter{equation}{0} \def\theequation{1.\arabic{equation}}

\bigskip

Сейчас (весна 2013 года) удачное время для обзорных лекций по
физике элементарных частиц. Предыдущий 2012 год называют годом
открытий -- на Большом Адронном Коллайдере ЦЕРНа (БАК, английское
название Large Hadron Collider, LHC) был найден бозон Хиггса;
также в ряде экспериментов был измерен один из углов смешивания
нейтрино $\theta_{13}$. Хотя смешивание нейтрино -- заметная часть
современной физики элементарных частиц, гораздо более важно
открытие бозона Хиггса и измерение его массы. Подтверждена
справедливость Стандартной Модели (СМ); следующий шаг -- открытие
Новой Физики. В начале 2013 года LHC встал на двухгодичный
перерыв, после которого его энергия будет увеличена в полтора
раза, что, быть может, приведет к открытию Новой Физики -- то есть
новых элементарных частиц за рамками СМ. Мы отложим до
заключительной шестой лекции обсуждение этих ожиданий, а первые
пять лекций будут посвящены изложению СМ, на сегодняшний день
успешно описывающей {\bf все} экспериментальные данные в области
физики элементарных частиц.

Предметом изучения физики элементарных частиц являются свойства
самых маленьких составляющих, из которых построены
(``составлены'') все окружающие нас предметы. В выходящем каждые
два года Обзоре Физики Частиц (Review of Particle Physics, RPP)
описаны свойства нескольких сотен частиц, что занимает около
полутора тысяч страниц текста, таблиц и рисунков в последнем
издании 2012 года. Большинство из этих частиц не являются
элементарными -- они состоят из более элементарных, или
фундаментальных, частиц. Поэтому удобно ввести понятие
фундаментальные частицы, из которых действительно всё составлено.
Фундаментальных частиц заметно меньше, чем элементарных, и их
свойства также описаны в RPP.

\newpage

\begin{center}

{\bf Таблица 1}

\bigskip

\begin{tabular}{|c|c|c|c|c|}
\hline
$s$ & 0 & 1/2 & 1 & 2 \\
\hline
& & $\nu_1 e$ ~~ $ud$ & $\gamma$ & \\
& $H$ & $\nu_2\mu$ ~~ $cs$ & $g$ & $h$ \\
& & $\nu_3 \tau$ ~~ $tb$ & $W^\pm$, $Z$ & \\
\hline
\end{tabular}
\end{center}

Все фундаментальные частицы перечислены в Таблице 1. Каждая
фундаментальная частица характеризуется своей массой, временем
жизни, спином и взаимодействиями, в которых она участвует.
Единственная известная сегодня фундаментальная частица со спином
ноль -- это бозон Хиггса, о котором уже говорилось. Массы всех
фундаментальных частиц появляются за счет механизма Хиггса, в
рамках которого и возникает эта частица со спином ноль. Более
подробное обсуждение механизма Хиггса будет дано в четвертой
лекции, посвященной электрослабым взаимодействиям.

Фундаментальные частицы со спином 1/2 бывают двух типов: лептоны и
кварки. Они объединяются в три кварк-лептонные семейства. Лептоны
первого семейства: нейтрино $\nu_1$ и электрон $e$. Кварки первого
семейства: $u$ (up, верхний) и $d$ (down, нижний). Второе
семейство: нейтрино $\nu_2$ и мюон $\mu$ и кварки $c$ (charm,
очарованный) и $s$ (strange, странный). Третье семейство: нейтрино
$\nu_3$ и тау-лептон $\tau$ и кварки $t$ (top) и $b$ (bottom).

Спин один имеет фотон ($\gamma$), глюоны ($g$) и промежуточные
бозоны слабых взаимодействий $W^\pm$ и $Z$. Здесь надо оговорить,
что глюоны являются переносчиками сильных взаимодействий, и их
имеется восемь штук аналогично тому, как каждый кварк может быть в
одном из трех цветовых состояний. Обмены глюонами связывают кварки
в адроны (сильно взаимодействующие частицы, в отличие от лептонов,
которые обладают только электромагнитными и слабыми
взаимодействиями). Большинство частиц, свойства которых описаны в
RPP, как раз и являются адронами, смотри посвященную сильным
взаимодействиям третью лекцию.

Наконец, имеется одна фундаментальная частица со спином 2,
гравитон $h$, обмен которой приводит к гравитационному притяжению.
Универсальность гравитации обусловлена взаимодействием гравитона с
тензором энергии-импульса элементарных частиц (а значит и с
тензором энергии-импульса состоящих из них макроскопических тел).

Мы перечислили все фундаментальные частицы и переходим к
обсуждению их масс. Фотон и гравитон безмассовы. Это следует из
дальнодействия гравитации и формы кулоновского потенциала. В обоих
случаях потенциал падает с расстоянием как $1/r$. Если бы обмен
осуществлялся массивной частицей с массой $\mu$, то взаимодействие
с расстоянием убывало бы гораздо быстрее, как $e^{-\mu r}/r$
(потенциал Юкавы). Глюоны также безмассовы, хотя сильные
взаимодействия характеризуются конечным радиусом порядка одного
ферми (1 fm$\equiv 10^{-13}$ см), что связано с обсуждаемым в
третьей лекции явлением конфайнмента, или невылетания цвета.

Все остальные фундаментальные частицы массивны. Перед тем, как
обсуждать величины их масс, следует сделать отступление о
размерностях, используемых в физике элементарных частиц. Удобной
является система единиц, в которой скорость света $c$ и постоянная
Планка $\hbar$ равны единице. Это значительно упрощает вид формул,
в которых остается одна размерная величина -- масса, или энергия.
Релятивистский закон дисперсии для свободной частицы $E^2 = m^2
c^4 + p^2 c^2$ превращается в $E^2 = m^2 + p^2$; энергия
покоящейся частицы равна ее массе. Это дает возможность измерять
массу частицы в единицах энергии. Удобной единицей энергии
является электронвольт, сокращенно эВ, или eV в английской
транскрипции. Соответствие с принятыми в макроскопической физике
единицами следующее:
\begin{equation}
1 \; \mbox {\rm эВ} \approx 1.6 \cdot 10^{-12} \; \mbox{\rm эрг
(СГСЭ)} = 1.6 \cdot 10^{-19} \; \mbox{\rm Дж (СИ)} \;\; .
\label{1.1}
\end{equation}
Используются следующие сокращения: $10^3$ эВ $\equiv 1$ кэВ;
$10^6$ эВ $\equiv$ 1 МэВ, $10^9$ эВ $\equiv 1$ ГэВ, $10^{12}$ эВ
$\equiv 1$ ТэВ. Масса электрона равна примерно 0.5 МэВ, поэтому
при двухквантовой аннигиляции находящихся в покое электрона и
позитрона $e^+ e^- \to 2\gamma$ энергия каждого фотона равна $0.5$
МэВ $\equiv 500$ кэВ. Масса протона $m_p = 938$ МэВ $\approx 1$
ГэВ $\equiv 10^9$ эВ выражается в граммах следующим образом:
\begin{equation}
m_p = \frac{E_0}{c^2} = \frac{10^9 \cdot 1.6 \cdot 10^{-12} \;
\mbox{\rm эрг}}{(3\cdot 10^{10} \; \mbox{\rm см/сек})^2} =
\frac{\mbox{грамм}}{6\cdot 10^{23}} \;\; , \label{1.2}
\end{equation}
в полном соответствии с известным значением числа Авогадро.
Постоянная Планка $\hbar \equiv h/2\pi \approx 10^{-27}$
эрг$\cdot$сек; значение произведения $\hbar c \approx 3\cdot
10^{-17} \; \mbox{\rm эрг}\cdot \mbox{\rm см} \approx 2\cdot
10^{-5}$ эв$\cdot$см позволяет легко переводить значения энергии
(или массы) в величины обратной длины и наоборот. Скажем,
комптоновская длина волны протона $l_p \equiv 1/m_p \approx 2\cdot
10^{-14}$ ГэВ$\cdot$см/1 ГэВ $= 0.2 \; {\rm fm}$ по порядку
величины совпадает с радиусом сильных взаимодействий, а
характерная величина сильного сечения $\sigma_{pp} \sim l_p^2 = 40
\cdot 10^{-27} \; \mbox{\rm см}^2 \equiv 40 \; {\rm mb}$, где 1
барн $\equiv 10^{-24} \; \mbox{\rm см}^2$ -- часто используемая в
физике элементарных частиц единица сечения, mb означает миллибарн
или $10^{-3}$ барна.

Отвечающее массе протона время $\tau_P \sim 2\cdot 10^{-14} \;
\mbox{\rm см}/c \approx 0.7 \cdot 10^{-24}$ сек характеризует
время жизни частиц, распадающихся за счет сильного взаимодействия
($\rho$-мезон, $\omega$-мезон и т.д.).

Постоянная тонкой структуры $\alpha$ -- чрезвычайно важная
величина, определяющая интенсивность (или силу) электромагнитных
взаимодействий. Из записанного в системе СГСЭ закона Кулона,
$F=e^2/r^2$, находим размерность квадрата заряда: $[e^2]
=$эрг$\cdot$см. Подставляя в качестве $e$ заряд электрона, для
безразмерной константы $\alpha$ получим:
\begin{equation}
\alpha\equiv\frac{e^2}{\hbar c} = \frac{[4.8 \cdot 10^{-10}
esu]^2}{3\cdot 10^{-17} \; \mbox{эрг}\cdot\mbox{см}} \approx
\frac{1}{137} \;\; . \label{1.3}
\end{equation}
Название ``постоянная тонкой структуры'' связано с тем, что эта
величина определяет тонкую структуру атомных спектров.

Приведем значения масс фундаментальных частиц. Начнем с заряженных
лептонов:
\begin{equation}
m_e \approx 0.511 \; \mbox{\rm МэВ}\; , \;\; m_\mu \approx 105.7
\; \mbox{\rm МэВ}\; , \;\; m_\tau \approx 1.78 \; \mbox{\rm ГэВ}\;
. \label{1.4}
\end{equation}
Массы нейтрино не измерены; измерены разности квадратов масс
нейтрино, оказавшиеся много меньше (1 эВ)$^2$, и имеется
экспериментальное ограничение на массу электронного нейтрино,
которая не может превышать двух электронвольт. Поэтому можно
утверждать, что массы нейтрино находятся в следующем интервале:
\begin{equation}
0\; \mbox{\rm эВ}\;\leq m_{\nu_1}, m_{\nu_2}, m_{\nu_3} \leq 2 \;
\mbox{эВ} \;\; . \label{1.5}
\end{equation}
Космология позволяет в несколько раз уменьшить верхнее ограничение
на массу нейтрино.

Что касается кварков, то их массы не могут быть измерены
непосредственно; из-за конфайнмента свободных кварков не
существует. Способы определения масс кварков изложены в третьей
лекции. Результаты таковы:
\begin{eqnarray}
m_u & \approx & 2\div 3 \mbox{\rm МэВ'а} \; , \;\; m_d \approx 5
\mbox{\rm МэВ} \; , \;\; m_s \approx 100 \mbox{\rm МэВ} \nonumber
\\
m_c & \approx & 1.3 \mbox{\rm ГэВ'а} \; , \;\; m_b \approx 4.5
\mbox{\rm ГэВ} \; , \;\; m_t = 174 \pm 1 \mbox{\rm ГэВ} \;\; .
\label{1.6}
\end{eqnarray}
Наконец, приведем массы промежуточных векторных бозонов и бозона
Хиггса:
\begin{equation}
M_{W^\pm} = 80.385 \pm 0.015 \; \mbox{\rm ГэВ} \; , \;\; M_Z =
91.188 \pm 0.002 \; \mbox{\rm ГэВ} \; , \;\; M_H = 126 \pm 1 \;
\mbox{\rm ГэВ} \;\; . \label{1.7}
\end{equation}

Стандартная Модель была сформулирована в конце 60-х -- начале 70-х
годов и всесторонне исследуется уже около сорока лет; свое
окончательное экспериментальное подтверждение она получила с
открытием в 2012 году бозона Хиггса.

Большинство фундаментальных частиц нестабильны: они распадаются на
более легкие частицы, если эти распады не запрещены законами
сохранения. При этом масса частицы оказывается комплексной: $M =
M_0 - i\Gamma/2$, где $\Gamma$ -- ширина (обратное время жизни)
частицы.

Безмассовые частицы ($\gamma$, $g$, $h$) стабильны -- их распады
запрещены кинематически. Также стабильно наиболее легкое нейтрино
-- его спин равен 1/2, и оно не может распадаться на безмассовые
бозоны. Более тяжелые нейтрино могут распадаться на легкие с
испусканием фотона: $\nu_i \to \nu_j \gamma$, однако вероятность
таких распадов исключительно мала, время жизни нейтрино гораздо
больше времени существования Вселенной, поэтому нейтрино могут
считаться стабильными частицами.

Кинематически разрешен распад электрона на нейтрино и фотон, либо
на три нейтрино. Эти распады запрещены законом сохранения
электрического заряда: будучи легчайшей заряженной частицей,
электрон стабилен. Более тяжелые заряженные лептоны распадаются:
мюон по каналу $\mu\to e \nu\bar\nu$, у $\tau$-лептона есть много
каналов распада. Среди составленных из кварков сильно
взаимодействующих частиц (адронов) стабильным является только
протон, распады которого запрещены сохранением барионного числа.

Из рассматриваемых в RPP элементарных частиц очень немногие
встречаются в природе (фотоны, электроны, протоны, входящие в
состав ядер нейтроны, а также нейтрино, детектировать которые
очень трудно). Остальные создаются в лабораториях на ускорителях,
либо рождаются при взаимодействии космических лучей с атмосферой
земли. В Большом Адронном Коллайдере протоны ускорялись в 2012
году до энергии 4 ТэВ'а, при этом суммарная энергия сталкивающихся
протонов в системе центра масс равнялась 8 ТэВ'ам. После
двухлетнего перерыва в 2015 году энергия протонов повысится до 7.5
ТэВ в каждом пучке. Найдем, чему такая суммарная энергия (а именно
она определяет максимальную массу рождаемых частиц) отвечает в
случае ускорителя с фиксированной мишенью:
\begin{equation}
(p_1 + p_2)^2 = 4 E^2_{\mbox{\rm ц.м.}} = 2m_p^2 + 2m_p
E_{\mbox{\rm л.с.}} \;\; , \label{1.8}
\end{equation}
и, пренебрегая массой протона, найдем:
\begin{equation}
E_{\mbox{\rm л.с.}} = 2 E_{\mbox{\rm ц.м.}} \cdot
\frac{E_{\mbox{\rm ц.м.}}}{m_p} \approx 10^5 \; \mbox{\rm ТэВ}
\equiv 10^{17} \; \mbox{\rm эВ} \;\; . \label{1.9}
\end{equation}

Преимущество коллайдеров (ускорителей со встречными пучками) для
рождения наиболее тяжелых частиц очевидно: ускорить протоны до
энергии $10^5$ ТэВ удастся не скоро. Именно поэтому наиболее
тяжелые фундаментальные частицы ($W^\pm$, $Z$, $t$ и, наконец,
$H$) были открыты в экспериментах на коллайдерах.

Закончим эту вводную лекцию обсуждением четырех типов
взаимодействий, в которых участвуют элементарные частицы.

Электромагнитные взаимодействия держат электроны в атомах; они же
ответственны за излучение фотона при переходах электронов в
атомах, скажем $2p \to 1 s$ $\gamma$. Взаимодействие
фундаментальных частиц с фотоном диагонально: испуская фотон,
заряженная частица остается собой. Для электрона амплитуда
излучения фотона пропорциональна $e(\bar e \gamma_\mu e)A_\mu$,
где $e$ -- заряд электрона, и в системе единиц $\hbar = c = 1$
имеем $e^2 = \alpha$.

Квантовая электродинамика основана на локальной абелевой $U(1)$
симметрии, и более подробно мы поговорим о ней во второй лекции.

Сильные взаимодействия удерживают нейтроны и протоны в ядрах. На
языке фундаментальных частиц сильное взаимодействие описывается
локальной $SU(3)$ калибровочной теорией: кварки испускают глюоны,
меняя свой цвет. Мы рассмотрим сильное взаимодействие в третьей
лекции.

Слабые взаимодействия ответственны за распады частиц: $n\to p e^-
\bar\nu_e$, $\mu\to e\nu_\mu\bar\nu_e$, $\tau \to
\mu\nu_\tau\bar\nu_\mu$ и т.д. Наряду с недиагональными
процессами, обусловленными излучением виртуального $W$-бозона,
имеются и диагональные слабые процессы, вызванные излучением
нейтрального $Z$-бозона. Основанная на локальной симметрии
$SU(2)_L \times U(1)$ электрослабая теория (объединяющая
электромагнитные и слабые взаимодействия) рассматривается в
четвертой лекции.

Наконец, играющая такую большую роль для макроскопических объектов
гравитация настолько слаба, что за исключением редчайших случаев в
физике элементарных частиц никакой роли не играет.

Квантовая теория поля -- это математический аппарат, необходимый
как для чтения и написания теоретических работ по физике
элементарных частиц, так и для понимания этой области физики. Наши
лекции -- облегченный обзор предмета; если читатель захочет
разобраться в нем подробнее, то начинать следует с изучения
квантовой теории поля по одной из следующих монографий: Ахиезер,
Берестецкий, ``Квантовая электродинамика''; Берестецкий, Лифшиц,
Питаевский, ``Квантовая электродинамика'', 4-й том курса Ландау и
Лифшица; Боголюбов, Ширков, ``Введение в теорию квантованных
полей''; Бьёркен, Дрелл, ``Релятивистская квантовая теория'';
Пескин, Шредер, ``Введение в квантовую теорию поля''.

В последующих лекциях будет более подробно излагаться Стандартная
Модель физики элементарных частиц, основанная на локальной
калибровочной $SU(3)_c \times SU(2)_L \times U(1)$ теории.

\newpage

\begin{center}

{\bf Лекция 2}

{\bf U(1) -- квантовая электродинамика (QED)}

\end{center}

\setcounter{equation}{0} \def\theequation{2.\arabic{equation}}

\bigskip

Начнем изложение с теории вещественного скалярного поля. Это
теория нейтральных частиц со спином ноль и, в частности, бозона
Хиггса. Свободно распространяющаяся плоская волна описывается
следующей формулой:
\begin{equation}
\varphi(x,t) = e^{-i p_\mu x_\mu} = e^{-i Et + i\bar p \bar x}
\;\; , \label{2.1}
\end{equation}
и отвечает свободной частице с четырехимпульсом $p_\mu = (E, \bar
p)$. Она является решением уравнения Клейна--Гордона:
\begin{equation}
(\Box + m^2) \varphi(x, t) = 0 \;\; , \label{2.2}
\end{equation}
где оператор Д'Аламбера $\Box \equiv \partial_\mu^2 \equiv
-(p_\mu)^2$, $p_\mu = i\partial_\mu = (i\frac{\partial}{\partial
t}, -i\frac{\partial}{\partial\bar x})$, и одним и тем же символом
$p_\mu$ обозначен как оператор 4-импульса, так и его собственное
значение.

Уравнение Клейна--Гордона следует из принципа наименьшего действия
$S = \int{\cal L} d^4 x$ с плотностью лагранжиана
\begin{equation}
{\cal L} = \frac{1}{2}(\partial_\mu\varphi)^2 - \frac{1}{2} m^2
\varphi^2 \;\; , \label{2.3}
\end{equation}
уравнение Эйлера--Лагранжа для которой
\begin{equation}
\frac{\delta{\cal L}}{\delta\varphi} = \frac{\partial}{\partial
x_\mu} \frac{\delta{\cal L}}{\delta\partial_\mu\varphi} \; , \;\;
\partial_\mu\varphi \equiv \frac{\partial\varphi}{\partial x_\mu}
\label{2.4}
\end{equation}
совпадает с (\ref{2.2}). В дальнейшем для краткости плотность
лагранжиана будем называть лагранжианом. В инвариантной
относительно преобразований Лоренца теории лагранжиан
Лоренц-инвариантен.

Лагранжиан свободного комплексного скалярного поля имеет следующий
вид:
\begin{equation}
{\cal L} = |\partial_\mu \Phi|^2 - m^2 \Phi^+ \Phi =\partial_\mu
\Phi^+ \partial_\mu \Phi - m^2 \Phi^+ \Phi \;\; , \label{2.5}
\end{equation}
уравнение движения совпадает с (\ref{2.2}) с заменой
$\varphi(x,t)$ на $\Phi(x,t)$. Комплексное поле описывает
заряженные частицы. Взаимодействие с электромагнитным полем
вводится путем ``удлинения'' производной:
\begin{equation}
p_\mu \to p_\mu - e A_\mu \; , \;\; i\partial_\mu \to
i\partial_\mu - eA_\mu \;\; , \label{2.6}
\end{equation}
\begin{equation}
{\cal L} = |(\partial_\mu + ie A_\mu)\Phi|^2 - m^2 \Phi^+ \Phi -
\frac{1}{4} F_{\mu\nu}^2 \;\; , \label{2.7}
\end{equation}
и в (\ref{2.7}) мы добавили лагранжиан электромагнитного поля
($F_{\mu\nu} \equiv \partial_\mu A_\nu - \partial_\nu A_\mu$).
Построенный таким образом лагранжиан инвариантен относительно
локальных (калибровочных) $U(1)$-преобразований:
\begin{equation}
\Phi = e^{i\lambda} \Phi^\prime \; , \;\; A_\mu = A_\mu^\prime -
\frac{1}{e} \partial_\mu \lambda \;\; , \label{2.8}
\end{equation}
где $\lambda$ зависит от координаты $x_\mu$ (если $\lambda$
одинаково во всем пространстве и не зависит от времени, то
преобразование называется глобальным; относительно таких
преобразований инвариантен лагранжиан (\ref{2.5}) -- введения поля
$A_\mu$ для глобальной симметрии не требуется). Записанный через
поля $A_\mu^\prime$ и $\Phi^\prime$ лагранжиан (\ref{2.7}) имеет
тот же вид, что и записанный в терминах полей $A_\mu$ и $\Phi$.

Перепишем (\ref{2.7}) в следующем виде:
\begin{equation}
{\cal L} = |\partial_\mu \Phi|^2 - m^2 \Phi^+ \Phi - \frac{1}{4}
F_{\mu\nu}^2 - eA_\mu[\Phi^+ i(\partial_\mu \Phi) -i(\partial_\mu
\Phi^+)\Phi] +e^2 A_\mu^2 \Phi^+ \Phi \;\; . \label{2.9}
\end{equation}
Здесь первые три члена описывают свободные поля $A_\mu$ и $\Phi$,
остальные члены описывают взаимодействие между ними. В качестве
поля $\Phi$ можно рассмотреть $\pi^+$-мезон: частицу со спином
$s=0$ и электрическим зарядом $+e$ (заряд электрона $-e$).
Комптон-эффект на $\pi^+$-мезоне (рассеяние фотона $\gamma$ на
$\pi^+$-мезоне) описывается в низшем порядке теории возмущений (в
порядке $e^2$ по константе связи) тремя диаграммами Фейнмана,
показанными на рис. 2.1.

\bigskip

\begin{center}
\bigskip
\includegraphics[width=1.\textwidth]{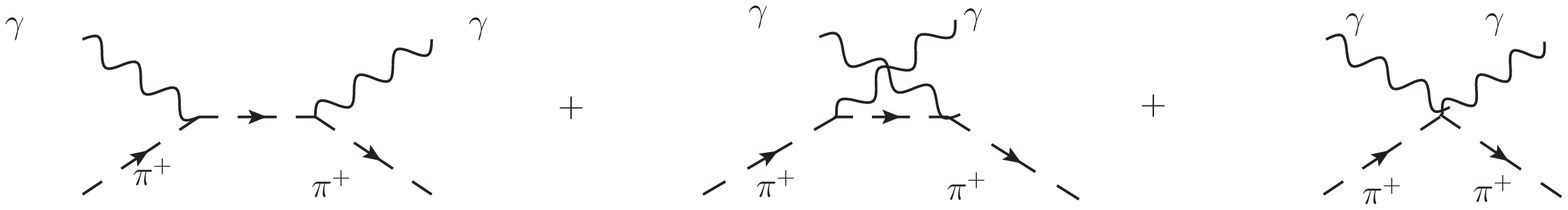}

\vspace{5mm}

{\it Рис. 2.1. Комптон-эффект на $\pi$-мезоне}

\end{center}

\bigskip

Первые две диаграммы возникают во втором порядке теории возмущений
по кубичному по полям члену в (\ref{2.9}), третья диаграмма -- в
первом порядке по четвертичному по полям последнему члену в
(\ref{2.9}). По диаграммам рис. 2.1 с помощью правил Фейнмана (или
правил диаграммной техники) выписывается амплитуда вероятности
реакции упругого рассеяния фотона на $\pi^+$-мезоне, $\gamma \pi^+
\to \gamma \pi^+$. Эта же амплитуда описывает процесс рождения
пары $\pi^+ \pi^-$ в столкновении двух фотонов: $\gamma\gamma \to
\pi^+ \pi^-$ и двухквантовую аннигиляцию $\pi^+ \pi^- \to
\gamma\gamma$. Зная амплитуду, можно найти сечение процесса
$\sigma$, которое по формуле $l=1/(n\sigma)$ позволяет определить
длину свободного пробега частицы в веществе. Пропорциональное
вероятности процесса сечение квадратично по амплитуде. Однако для
рассматриваемых реакций большее практическое значение имеет
вычисление числа рождаемых $\pi^+ \pi^-$-пар на
$\gamma\gamma$-коллайдере согласно формуле $N=\sigma \cdot L$, где
$L$ -- светимость коллайдера ($[N] = 1/\mbox{\rm сек} \; , \;\;
[\sigma] = \mbox{\rm см}^2 \; , \;\; [L] = 1/(\mbox{\rm сек}\cdot
\mbox{\rm см}^2)$).

Перейдем к электромагнитному взаимодействию частиц со спином $1/2$
(заряженные лептоны и кварки Стандартной Модели). Соответствующее
поле описывается биспинором $\psi$, являющимся решением уравнения
Дирака:
\begin{equation}
(p_\mu \gamma_\mu - m) \psi = 0 \;\; , \label{2.10}
\end{equation}
где $p_\mu \gamma_\mu \equiv \hat p = p_0 \gamma_0 - p_i
\gamma_i$. Умножая (\ref{2.10}) слева на $\hat p +m$ и требуя
выполнения дисперсионного соотношения для свободной частицы $p^2 -
m^2 =0$, придем к основному свойству матриц Дирака $\gamma_\mu$:
$\gamma_\mu \gamma_\nu + \gamma_\nu \gamma_\mu = 2g_{\mu\nu}$.
Удовлетворяющие этому соотношению матрицы $4\times 4$ в
стандартном представлении имеют следующий вид:
\begin{equation}
\gamma_0 = \left(
\begin{array}{rr}
1 & 0 \\
0 & -1
\end{array}
\right) \; , \;\; \gamma_i = \left(
\begin{array}{cc}
0 & \sigma_i \\
-\sigma_i & 0
\end{array}
\right) \;\; . \label{2.11}
\end{equation}
Здесь $\sigma_i$ -- матрицы Паули:
\begin{equation}
\sigma_1 = \left(
\begin{array}{cc}
0 & 1 \\
1 & 0
\end{array}
\right) \; , \;\; \sigma_2 = \left(
\begin{array}{rr}
0 & -i \\
i & 0
\end{array}
\right)\; ,  \;\; \sigma_3 = \left(
\begin{array}{rr}
1 & 0 \\
0 & -1
\end{array}
\right). \label{2.12}
\end{equation}

Подставляя в уравнение (\ref{2.10}) биспинор $\psi$ в виде $\psi =
\left(\begin{array}{c} \varphi \\ \chi \end{array}\right)$, где
$\varphi$ и $\chi$ -- двухкомпонентные спиноры, получим:
\begin{equation}
\left(
\begin{array}{cc}
E-m & -\bar p \; \bar\sigma \\
\bar p \; \bar\sigma & -E-m
\end{array}
\right) \left(
\begin{array}{c}
\varphi \\
\chi
\end{array}
\right) = 0 \;\; , \label{2.13}
\end{equation}

В нерелятивистском пределе $\bar p \to 0$ спинор $\chi$ обращается
в ноль. Остается двухкомпонентный спинор $\varphi$, описывающий
нерелятивистскую частицу со спином $1/2$.

Взаимодействие с электромагнитным полем вводится ``удлинением''
производной, заменяющим свободное уравнение Дирака (\ref{2.10}) на
\begin{equation}
(\hat p - e \hat A - m)\psi = 0 \;\; , \label{2.14}
\end{equation}
где $\hat A \equiv A_\mu \gamma_\mu$. В качестве упражнения
полезно вывести из (\ref{2.14}) в нерелятивистском пределе
взаимодействие спина электрона с магнитным полем. Оно отвечает
магнитному моменту электрона $\vec\mu = 2\cdot (e/2m_e)\vec s$,
или гиромагнитному отношению $g_e = 2$ (в древесном приближении),
в то время как для орбитального движения магнитный момент равен
$(e/2m_e)\vec l$ и гиромагнитное отношение равно единице.

Лагранжиан дираковского фермиона со спином $1/2$ и единичным
электрическим зарядом, взаимодействующего с электромагнитным
полем, имеет следующий вид:
\begin{equation}
{\cal L} = \bar\psi(\hat p - e\hat A -m)\psi-
\frac{1}{4}F_{\mu\nu}^2 = \bar\psi(\hat p -m)\psi -
\frac{1}{4}F_{\mu\nu}^2 - e\bar\psi\hat A\psi \;\; . \label{2.15}
\end{equation}
Взаимодействие описывается членом $e\bar\psi A\psi$; на рис. 2.2
приведены диаграммы, описывающие в первом неисчезающем приближении
теории возмущений Комптон-эффект на электроне.

\begin{center}
\bigskip
\includegraphics[width=1.\textwidth]{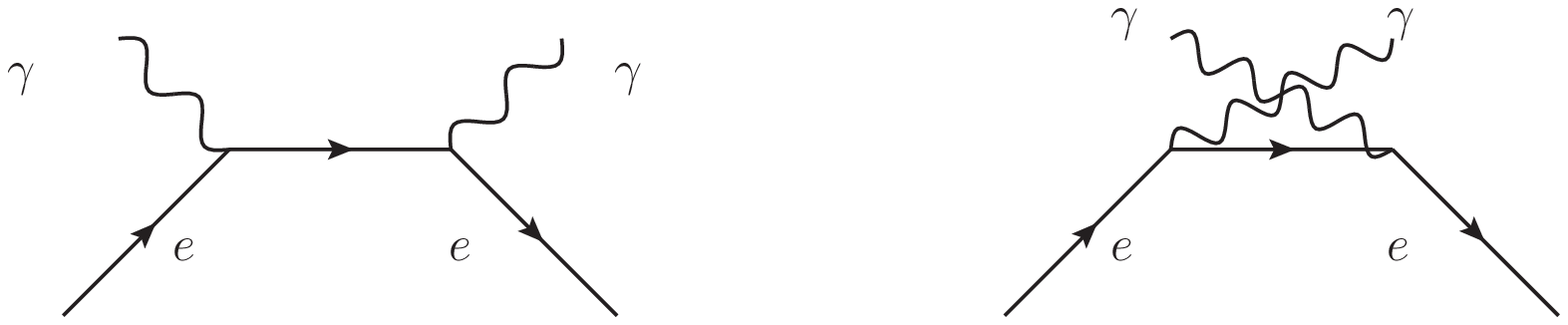}

\vspace{5mm}

{\it Рис. 2.2. Комптон-эффект на электроне}

\end{center}

\bigskip

Другой важный в физике элементарных частиц процесс -- рождение
пары $\mu^+ \mu^-$ в $e^+ e^-$-аннигиляции -- описывается
приведенной на рис. 2.3 диаграммой. При инвариантной энергии
$\sqrt s = \sqrt{(p_{e^+} + p_{e^-})^2} = 2 E_{\mbox{\rm ц.м.}}
\gg 2m_\mu$ сечение этого процесса дается следующей формулой:
\begin{equation}
\sigma = \frac{4\pi}{3} \frac{\alpha^2}{s} \;\; . \label{2.16}
\end{equation}

\begin{center}
\bigskip
\includegraphics[width=.6\textwidth]{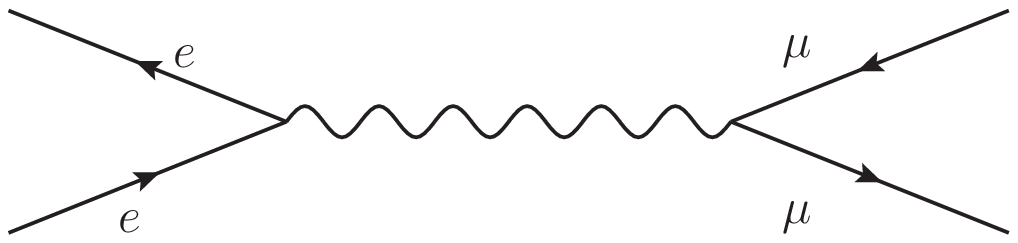}

\vspace{5mm}

{\it Рис. 2.3. Рождение пары $\mu^+ \mu^-$ в $e^+
e^-$-аннигиляции}

\end{center}

\bigskip

Пропорциональность сечения $1/s$ следует из соображений
размерности (в рассматриваемом случае высоких энергий можно
пренебречь как массой электрона, так и массой мюона);
пропорциональность $\alpha^2 = e^4$ следует из того факта, что в
каждой вершине рис. 2.3 стоит заряд электрона $e$, а сечение
пропорционально квадрату вычисляемой по этой диаграмме амплитуды.
Поэтому аккуратное вычисление сечения необходимо только для
нахождения численного множителя $4\pi/3$.

Как уже было сказано, в древесном приближении гиромагнитное
отношение для спина электрона, следующее из уравнения Дирака,
равно двум. Изображенные на рис. 2.4 петлевые поправки изменяют
древесное значение. В одной петле поправка была вычислена
Швингером на заре квантовой электродинамики. Она оказалась равной
$\alpha/2\pi \approx 0.00116$. К настоящему времени вычислены все
графики до четырех петель включительно, и сравнение с
экспериментальным результатом для величины электронной аномалии
$a_e \equiv (g_e -2)/2$ позволяет получить наиболее точное
значение постоянной тонкой структуры $\alpha^{-1} =
137.035999084(51)$. Оно используется для получения теоретического
предсказания значения мюонной аномалии:
\begin{equation}
a_\mu^{\mbox{\rm теор}} \equiv \left(\frac{g_\mu
-2}{2}\right)^{\mbox{\rm теор}} = [1165.9183(5)] \cdot 10^{-6}
\;\; , \label{2.17}
\end{equation}
в то время как экспериментальное значение
\begin{equation}
a_\mu^{\mbox{\rm эксп}} = [1165.9209(6)] \cdot 10^{-6} \;\; .
\label{2.18}
\end{equation}
Неопределенности в последних значащих цифрах приведены в круглых
скобках; отличие экспериментального числа от теоретического
находится на уровне трех стандартных отклонений, что (быть может)
свидетельствует о наличии Новой Физики. Для ее поиска
предпочтительна именно мюонная аномалия: вклад тяжелых частиц с
массой $M$ подавлен как $m_l^2/M^2$, поэтому в $a_\mu$ он гораздо
больше, чем в $a_e$.

\begin{center}
\bigskip
\includegraphics[width=.6\textwidth]{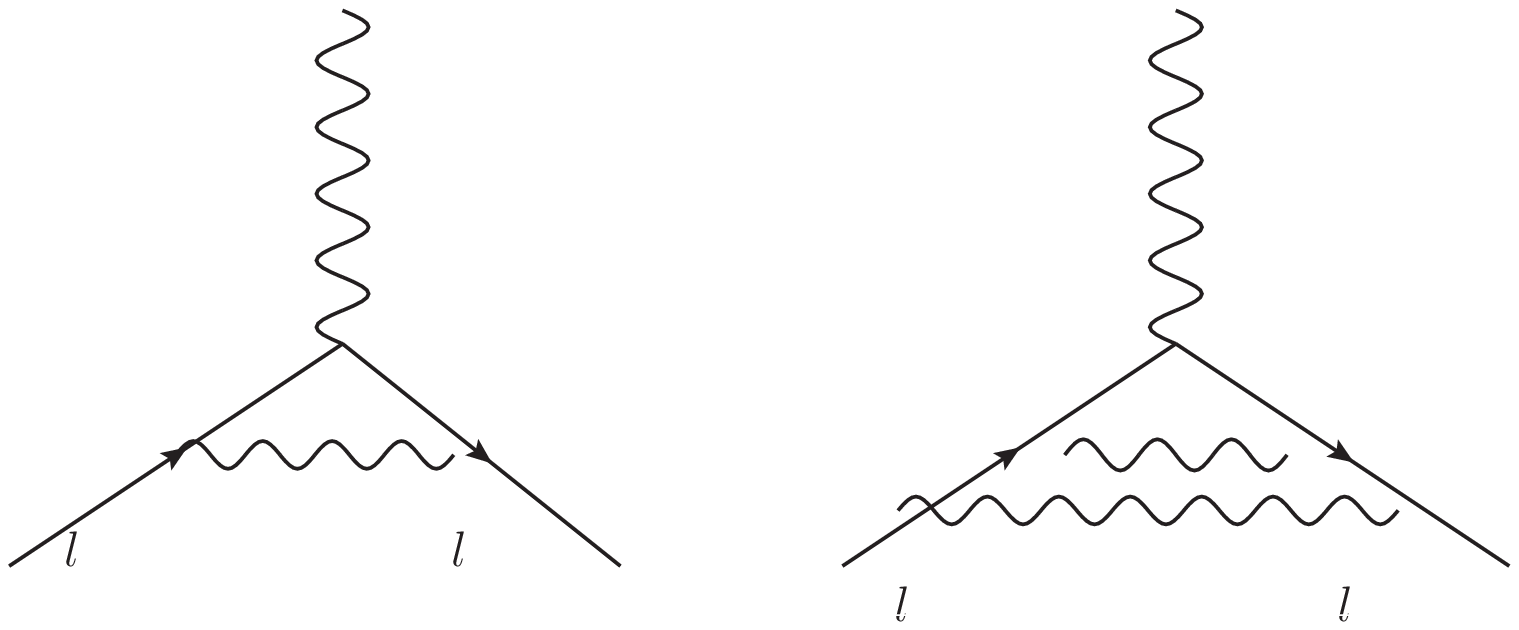}

\vspace{5mm}

{\it Рис. 2.4. Аномальный магнитный момент лептона $l$}

\end{center}

Обсудим некоторые свойства позитрония -- связанного состояния
электрона и позитрона. Кулоновское притяжение приводит к появлению
связанных состояний со спектром $E_n = -\mu e^4/(2n^2)$, где
$n=1,2,...$, а $\mu = m_e/2$ -- приведенная масса электрона и
позитрона. Основному состоянию отвечает $n=1$, $L=0$, а полный
момент $J=1$ для ортопозитрония и $J=0$ для парапозитрония и
получается сложением спинов $e^+$ и $e^-$. Будучи нейтральной
системой, позитроний обладает определенной зарядовой четностью
$C$. Под действием зарядового сопряжения электрон переходит в
позитрон, а позитрон -- в электрон. Волновая функция электрона и
позитрона меняет знак при их перестановке, как и при перестановке
любых двух фермионов. Учитывая поведение координатной и спиновой
волновых функций при перестановке, получаем:
\begin{equation}
C(e^+ e^-) = (e^- e^+) = -(-1)^L (-1)^{S+1} (e^+ e^-) \; , \;\; C=
(-1)^{L+S} \;\; , \label{2.19}
\end{equation}
где $S$ - сумма спинов электрона и позитрона.

Основное состояние парапозитрония $C$-четно, а ортопозитрония --
$C$-нечетно.

Аннигиляция пары $e^+ e^-$ приводит к распаду позитрония.
Наибольшую вероятность имеет двухфотонная аннигиляция $e^+ e^- \to
2\gamma$ (однофотонная аннигиляция запрещена кинематически, для
реального фотона $q^2 =0$). Отрицательная зарядовая четность
фотона ($C\bar E = -\bar E$, $C\bar H = -\bar H$) запрещает распад
основного состояния ортопозитрония на два фотона; его время жизни
определяется трехфотонной аннигиляцией, $\tau_{\mbox{\rm орто}} =
1.4 \cdot 10^{-7}$ сек, в то время как для распадающегося на два
фотона парапозитрония $\tau_{\mbox{\rm пара}} = 1.2 \cdot 10
^{-10}$ сек. Лишний множитель $\alpha$ в вероятности распада
обуславливает такое большое различие во временах жизни орто- и
парапозитрониев. Электромагнитное взаимодействие электронов
записывается в виде $j_\mu A_\mu$; при $C$-преобразовании
электроны заменяются на позитроны, поэтому $C j_\mu = -j_\mu$, и
$C$-инвариантность КЭД требует $CA_\mu = -A_\mu$ -- еще один
способ понять отрицательную $C$-четность фотона.

Обсудим пространственную четность позитрония. При зеркальном
отражении $\vec x \to -\vec x$ пространственная часть волновой
функции умножается на $(-1)^L$, но оказывается, что это не всё. В
релятивистской теории, изучающей рождения и распады частиц, важно
понятие внутренней $P$-четности частицы. Внутренние четности
электрона и позитрона противоположны, что является фундаментальным
свойством уравнения Дирака. Поэтому $P$-четность позитрония равна
$(-1)^{L+1}$, она отрицательна у основных состояний орто- и
парапозитрониев.

Обозначая квантовые числа позитрония в виде $J^{PC}$, находим для
основного состояния парапозитрония $J^{PC} = 0^{-+}$, для
ортопозитрония $J^{PC} = 1^{--}$.

Сильно взаимодействующие частицы с целым спином называются
мезонами; они состоят из пары кварк - антикварк, и их квантовые
числа определяются теми же формулами, что и полученные нами для
позитрония. Легчайшими адронами (общее название сильно
взаимодействующих частиц) являются $\pi$-мезоны. $\pi^0$-мезон
``состоит'' из $u\bar u$- и $d\bar d$-кварков и имеет спин ноль.
Поэтому аналогично парапозитронию квантовые числа $\pi^0$-мезона
равны $0^{-+}$, и он распадается на два фотона. В системе покоя
$\pi^0$-мезона эти фотоны летят в противоположные стороны, и их
линейные поляризации перпендикулярны, так как псевдоскалярная
комбинация волновых функций двух фотонов имеет вид $F_1 \tilde F_2
= \frac{1}{2}\varepsilon_{\mu\nu\rho\sigma}F_{1_{\mu\nu}}
F_{2_{\rho\sigma}} \sim \varepsilon_{ikl}k_i E_k^1 E_l^2$, где
$k_i$ -- трехмерный импульс одного из фотонов. Если бы $\pi^0$ был
скаляром, то линейные поляризации фотонов были бы параллельны:
$F_1 F_2 \equiv F_{1_{\mu\nu}} F_{2_{\mu\nu}} \sim E_i^1 \cdot
E_k^2$. Псевдоскалярность $\pi^0$-мезона была проверена в 50-е
годы, на заре физики элементарных частиц. Масса $\pi^0$-мезона
равна 135 МэВ, и измерить поляризацию фотонов такой большой
энергии непосредственно затруднительно. Поэтому анализируются
распады, в которых оба фотона, будучи виртуальными, конвертируют в
пары $e^+ e^-$, т.е. распады $\pi^0 \to e^+ e^+ e^- e^-$. Эти
распады очень редки; на 100000 распадов $\pi^0 \to\gamma\gamma$
приходится примерно три распада $\pi^0 \to e^+ e^+ e^- e^-$.
Изучается корреляция плоскостей, в которых разлетаются пары $e^+
e^-$. Эти плоскости оказываются преимущественно перпендикулярными
друг другу, что свидетельствует о комбинации квантовых чисел
$\pi^0$-мезона $J^P = 0^-$. Аналогичное рассмотрение
использовалось в 2013 году для установления $P$-четности бозона
Хиггса. Распады $H\to \gamma\gamma$ были найдены, однако их
относительная вероятность мала, и распады $H\to\gamma\gamma \to
e^+ e^+ e^- e^-$ вряд ли будут когда-либо обнаружены. На помощь
приходят распады $H\to ZZ \to e^+ e^+ e^- e^-$. На пару
$Z$-бозонов распадается заметная доля бозонов Хиггса, а около 10\%
распадов $Z$ приходится на пару заряженных лептонов. Изучение
корреляции плоскостей разлета этих лептонов показало, что они, в
основном, параллельны, т.е. для бозона Хиггса $J^P = 0^+$, что
соответствует теоретическим ожиданиям. $H$ -- скалярный бозон.

В заключение этой главы обсудим зависимость интенсивности
электромагнитного взаимодействия от расстояния, или ``бег''
``константы'' $\alpha$. Равная $1/137...$ на больших расстояниях,
она растет, когда характерные расстояния становятся меньше, чем
$1/m_e$ (передаваемые импульсы больше, чем $m_e$). За это явление
ответственны приведенные на рис. 2.5 диаграммы.

\begin{center}
\bigskip
\includegraphics[width=.6\textwidth]{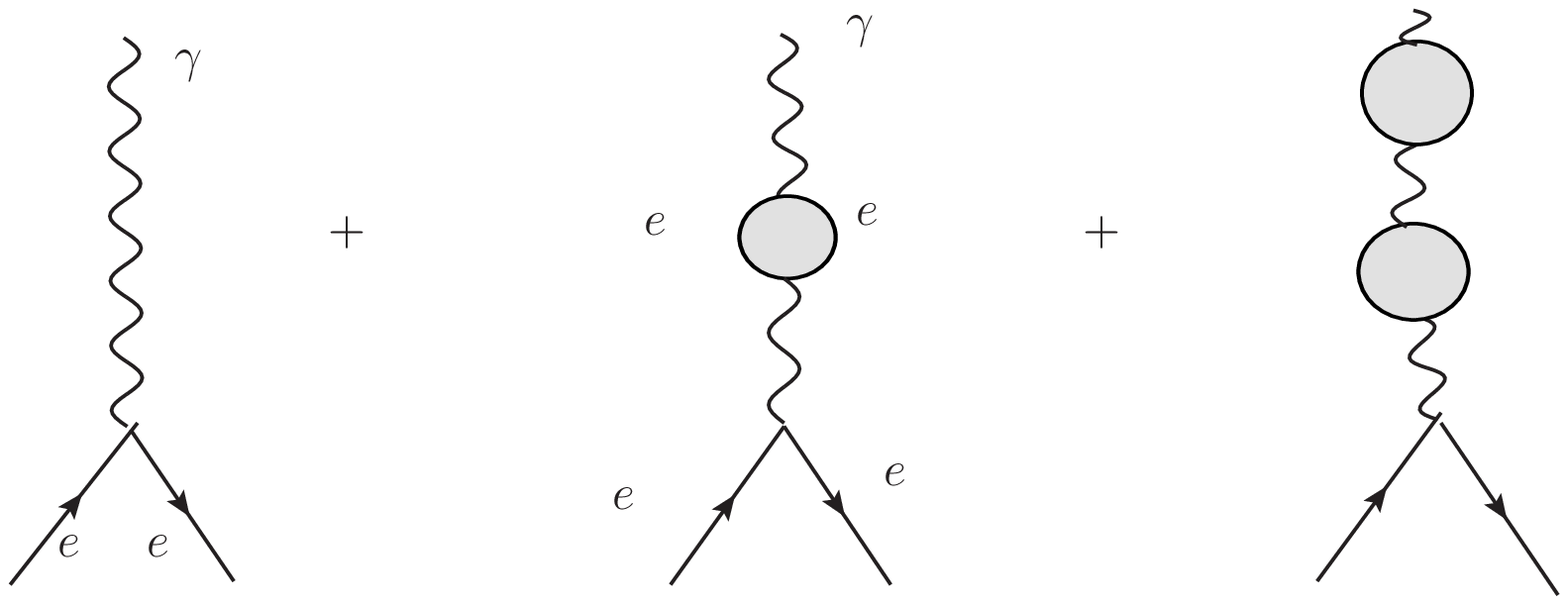}

\vspace{5mm}

{\it Рис. 2.5. Диаграммы, приводящие к бегу заряда в КЭД}

\end{center}

\bigskip

Их суммирование приводит к следующей знаменитой формуле:
\begin{equation}
\alpha(-q^2) = \frac{\alpha}{1-\frac{\alpha}{3\pi}
\ln\left(\frac{-q^2}{m_e^2}\right)} \;\; . \label{2.20}
\end{equation}

Учет $\mu^+ \mu^-$, $\tau^+ \tau^-$ и кварк-антикварковых пар
ведет к росту коэффициента перед логарифмом.

С ростом $-q^2$ (уменьшением расстояния) эффективный квадрат
заряда $\alpha(-q^2)$ растет, что соответствует экранировке заряда
на больших расстояниях рождаемыми в вакууме виртуальными $e^+
e^-$-парами. Явление роста заряда проверено экспериментально: в
физике промежуточных $W$- и $Z$-бозонов адекватной величиной
является $\alpha(M_Z^2) \approx 1/128$, на примерно 6\%
превосходящая постоянную тонкой структуры $\alpha = 1/137...$. На
экспоненциально малых расстояниях в (\ref{2.20}) имеется полюс; на
таких расстояниях заряд обращается в бесконечность. Можно говорить
о несамосогласованности КЭД как теории со слабым взаимодействием:
на малых расстояниях взаимодействия становятся сильными. Если
потребовать малости заряда на сколь угодно малых расстояниях, то
заряд на больших расстояниях обратится в ноль. Это явление,
открытое в 50-е годы прошлого века, получило название ноль-заряда
(или московский ноль). Оно оказалось присущим всем известным в то
время теориям поля в четырехмерном пространстве-времени.
Единственное исключение -- неабелевы калибровочные теории, в
которых имеет место антиэкранировка заряда, и с уменьшением
расстояния заряд уменьшается. Что касается КЭД, или $U(1)$ заряда
в Стандартной $SU(3)_c \times SU(2)_L \times U(1)$ теории, то
можно думать, что еще до ноль-зарядного полюса теория
видоизменяется, тем самым решая эту проблему.

\newpage

\begin{center}

{\bf Лекция 3}

{\bf $SU(3)_c$ -- квантовая хромодинамика (КХД, QCD)}

\end{center}

\setcounter{equation}{0} \def\theequation{3.\arabic{equation}}

\bigskip

В трех кварк-лептонных поколениях Стандартной Модели насчитывается
шесть кварков: $u$, $d$, $s$, $c$, $b$ и $t$, где каждая буква
означает Дираковский биспинор. Каждый из кварков может находиться
в одном из трех цветовых состояний:
\begin{equation}
q_i = \left(
\begin{array}{c}
q_1 \\
q_2 \\
q_3
\end{array}
\right) \;\; . \label{3.1}
\end{equation}

Динамику кварков описывает локальная цветовая $SU(3)_c$-симметрия,
$c$ -- от слова ``colour'', цвет. $SU(3)$ -- группа унитарных $3
\times 3$ матриц $S$ с равным единице детерминантом. Начнем
изложение с более простой группы глобальных преобразований
$SU(2)$, элемент которой имеет следующий вид:
\begin{equation}
S = e^{i \Lambda_i T_i} \;\; , \label{3.2}
\end{equation}
где $T_i \equiv 1/2 \sigma_i$ -- генераторы $SU(2)$, а $\Lambda_i$
-- три произвольных параметра. Из эрмитовости матриц Паули следует
унитарность матриц $S$. Равенство их детерминантов единице следует
из формулы
\begin{equation}
{\rm det} \exp(Z) = \exp ({\rm Tr} Z) \;\; . \label{3.3}
\end{equation}
Фундаментальное представление $SU(2)$ -- двухкомпонентный спинор.

Произвольный элемент группы $SU(3)$ также записывается в виде
(\ref{3.2}), но теперь $T_i \equiv 1/2\lambda_i$, где $\lambda_i$
-- восемь матриц Гелл-Манна:
\begin{equation}
\lambda_{1,2,3} = \left(
\begin{array}{cc}
& 0 \\
\sigma_{1,2,3} & 0 \\
0 ~~~ 0 & 0
\end{array}
\right) \;\; , \;\; \lambda^{4(5)} = \left(
\begin{array}{ccc}
0 & 0 & 1(-i) \\
0 & 0 & 0 \\
1(i) & 0 & 0
\end{array}
\right) \;\; , \label{3.4}
\end{equation}
$$
\lambda^{6(7)} = \left(
\begin{array}{ccc}
0 & 0 & 0 \\
0 & 0 & 1(-i)\\
0 & 1(i) & 0
\end{array}
\right) \;\; , \lambda^8 = \frac{1}{\sqrt 3} \left(
\begin{array}{ccc}
1 & 0 & 0 \\
0 & 1 & 0 \\
0 & 0 & -2
\end{array}
\right) \;\; .
$$

Если преобразование локальное, то его параметры $\Lambda_i$
различны в разных точках пространства-времени, $\Lambda_i \equiv
\Lambda_i(x)$.

Для инвариантности кинетического члена кваркового лагранжиана
относительно локальных $SU(3)_c$-преобразований кварковых полей
$q(x) = S(x) q^\prime(x)$ производную следует ``удлинить''
аналогично случаю $U(1)$- преобразований:
\begin{equation}
\bar q \gamma_\mu (\partial_\mu - i g A_\mu) q = \bar q^\prime
\gamma_\mu(\partial_\mu - i g A_\mu^\prime) q^\prime \;\; ,
\label{3.5}
\end{equation}
где $A_\mu \equiv A_\mu^i \frac{1}{2}\lambda_i$. При испускании
или поглощении глюона меняется ``цвет'' кварка, тогда как его
``тип'', или флэйвор, не меняется: $u$-кварк остается $u$-кварком,
$d$-кварк -- $d$-кварком и т.д.

Требование инвариантности (\ref{3.5}) дает закон преобразований
векторных полей $A_\mu(x)$:
\begin{equation}
-ig A_\mu = S^+ \partial_\mu S - ig S^+ A_\mu S \;\; , \label{3.6}
\end{equation}
$$
A_\mu = S^+ A_\mu^\prime S + \frac{i}{g} S^+ \partial_\mu S \;\; .
$$

Восьмерка глюонных полей $A_\mu^i$ преобразуется по
присоединенному представлению $SU(3)_c$ (первый член после знака
равенства в последней формуле), однако наряду с однородным имеется
и неоднородный член, аналогичный случаю $U(1)$-симметрии. В
отличие от фотонного поля, глюонные поля преобразуются и при
постоянных в пространстве-времени преобразованиях. Это означает,
что в отличие от нейтрального фотона ($Q_\gamma = 0$) глюоны имеют
цветовые заряды: глюоны излучают глюоны.

Понятие ковариантной производной $D_\mu$ позволяет получить
глюонный лагранжиан аналогично лагранжиану фотонов. В случае
группы $U(1)$
\begin{equation}
D_\mu \equiv \partial_\mu - i g A_\mu \;\; , \label{3.7}
\end{equation}
$$
F_{\mu\nu} = D_\mu A_\nu - D_\nu A_\mu = \partial_\mu A_\nu -
\partial_\nu A_\mu
$$
в силу того, что фотонные поля $A_\mu$ и $A_\nu$ коммутируют.
Тензор  электромагнитного поля $F_{\mu\nu}$ инвариантен
относительно локальных $U(1)$-преобразований; Лоренц-инвариантный
лагранжиан фотонов равен:
\begin{equation}
{\cal L} = -\frac{1}{4} F_{\mu\nu}^2 \;\; . \label{3.8}
\end{equation}

В случае глюонов поля $A_\mu \equiv A_\mu^i \frac{1}{2}\lambda_i$,
будучи матрицами, не коммутируют, поэтому
\begin{equation}
G_{\mu\nu} = D_\mu A_\nu - \partial_\nu A_\mu = \partial_\mu A_\nu
- \partial_\nu A_\mu - ig[A_\mu , A_\nu] \;\; . \label{3.9}
\end{equation}
Нетрудно найти закон преобразования тензора глюонного поля при
калибровочных преобразованиях (\ref{3.6}):
\begin{equation}
G_{\mu\nu} = S^+ G_{\mu\nu} S \;\; . \label{3.10}
\end{equation}
Инвариантный лагранжиан глюонного поля
\begin{equation}
{\cal L} = -\frac{1}{2} {\rm Tr} G_{\mu\nu}^2 = -\frac{1}{4}
G_{\mu\nu}^{i^2} \; , \;\; i = 1, 2, ..., 8 \label{3.11}
\end{equation}
наряду с линейными по полям членами содержит изображенные на рис.
3.1 тройную и четверную вершины -- неабелева локальная симметрия с
необходимостью приводит к взаимодействиям векторных полей друг с
другом.

\begin{center}
\bigskip
\includegraphics[width=.8\textwidth]{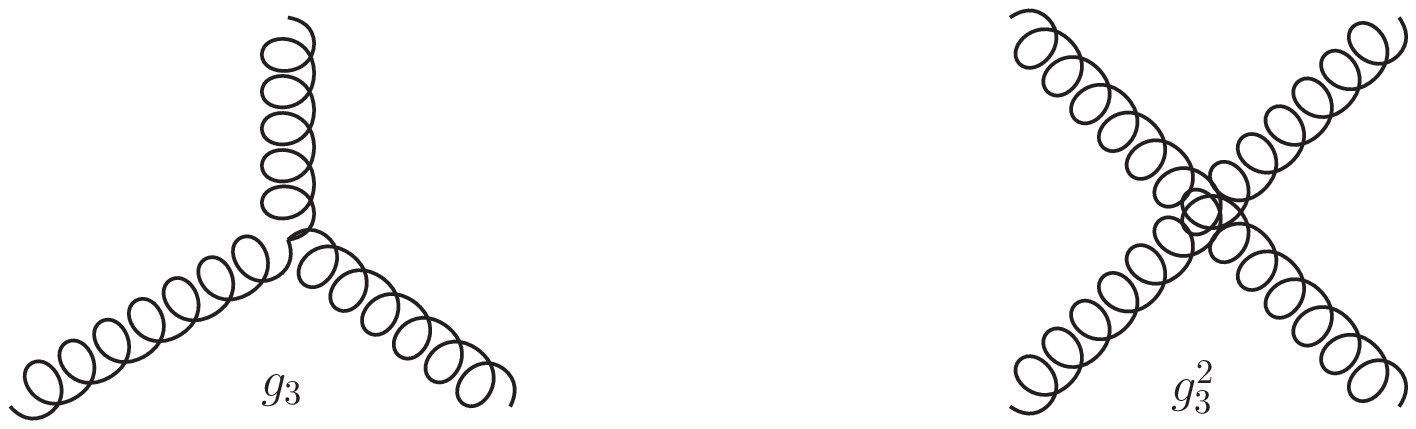}

\vspace{5mm}

{\it Рис. 3.1. Самодействие глюонов. Индекс ``3'' у зарядов
напоминает, что речь идет о группе $SU(3)_c$}

\end{center}

\bigskip

Обмены глюонами связывают кварки в адроны аналогично тому, как
обмены фотонами удерживают электрон и позитрон в позитронии. Но
имеется существенное отличие: самодействие глюонов приводит к
росту взаимодействия с увеличением расстояния, что, в свою
очередь, приводит к явлению конфайнмента (пленения) кварков:
свободные кварки не могут вылететь из адронов. Для случая
неабелевой цветовой симметрии приводящие к бегу заряда диаграммы
показаны на рис. 3.2. Сравнивая с рис. 2.5., мы видим
дополнительную диаграмму, обусловленную самодействием глюонов.

\begin{center}
\bigskip
\includegraphics[width=.8\textwidth]{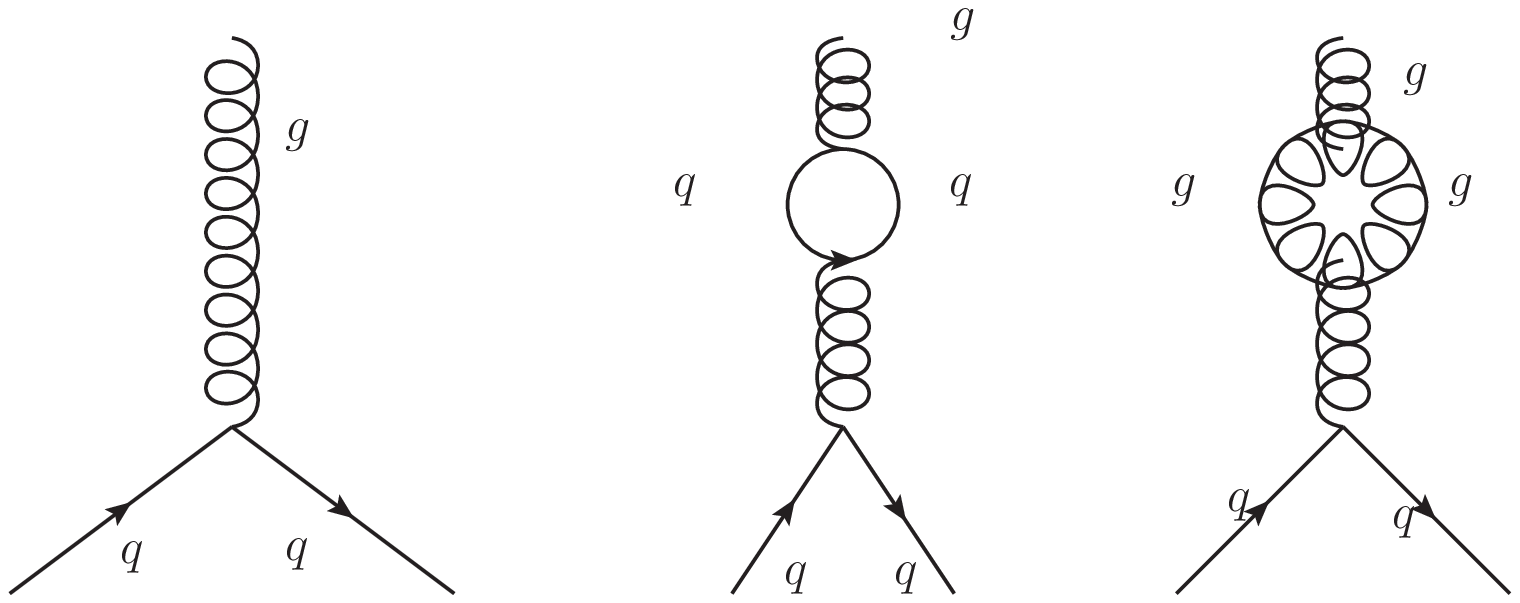}

\vspace{5mm}

{\it Рис. 3.2. Диаграммы, приводящие к зависимости цветового
заряда от расстояния}

\end{center}

\bigskip

Вычисление приводит к следующей зависимости:
\begin{eqnarray}
\alpha_3(-q^2) & = & \frac{\alpha_3(\mu^2)}{1+
\frac{\alpha_3(\mu^2)}{4\pi}(11-\frac{2}{3} N_q)
\ln\left(-\frac{q^2}{\mu^2}\right)} \equiv \nonumber \\
& \equiv & \frac{4\pi}{(11- \frac{2}{3} N_q)
\ln\left(-\frac{q^2}{\Lambda_{\rm QCD}^2}\right)} \;\; ,
\label{3.12}
\end{eqnarray}
где $N_q$ -- число типов кварков, масса которых меньше, чем
$\mu^2$ (то есть при $\mu = 200$ ГэВ имеем $N_q = 6$). Кварки с
массами больше $q^2$ в (\ref{3.12}) не учитываются; если же $-q^2
> m_q^2 > \mu^2$, то (\ref{3.12}) видоизменяется: от $\mu^2$ до
$m_q^2$ в $N_q$ такой кварк не учитывается, а от $m_q^2$ до $-q^2$
значение $N_q$ увеличивается на единицу.

Вклад, пропорциональный $-2/3 N_q$, обусловлен кварковой петлей, и
он совпадает с абелевым случаем (\ref{2.20}) с точностью до
фактора $1/2$, происходящего от ${\rm Tr} T_i^+ T_k = 1/2
\delta_{ik}$, $T_i \equiv 1/2 \lambda_i$ -- генератор $SU(3)$ в
фундаментальном представлении, стоящий в кварк-кварк-глюонной
вершине кварковой петли на рис. 3.2, смотри уравнение (\ref{3.5}).
Знак этого члена тот же, что и в абелевом случае, поэтому он
отвечает экранировке заряда. Имеющий противоположный знак член,
пропорциональный 11, возникает от глюонной петли. Численно он
превосходит вклад кварков, поэтому в КХД имеет место
антиэкранировка: с ростом расстояния между цветными объектами
взаимодействие растет. При этом методы теории возмущений по малой
константе связи, давшие так много замечательных результатов в КЭД,
становятся неприменимыми. В частности, мы не можем найти
аналитически спектр бесцветных связанных состояний кварков. Были
развиты численные методы, и они дают близкие к измеренным на
эксперименте значения масс адронов.

Как уже было сказано, из-за конфайнмента масса цветных объектов
бесконечна, конечные же массы имеют бесцветные (``белые'')
объекты. Скажем, мезоны, состоящие из пары кварк-актикварк.
Качественная картина конфайнмента такова: при ``растаскивании''
кварка и антикварка из мезона между ними натягивается глюонная
струна, энергия которой на единицу длины постоянна (она порядка
$\Lambda_{\rm QCD}^2$). В какой-то момент струна лопается, и на
месте разрыва образуется кварк-антикварковая пара. Физический
процесс -- распад (бесцветного) мезона на пару (бесцветных)
мезонов под действием внешнего возбуждения, скажем, фотона.

Отметим, что лагранжиан КХД не содержит параметра $\Lambda_{\rm
QCD}$. Он возникает из параметра $\mu$, который вводится для
инфракрасной регуляризации интеграла, отвечающего диаграмме
Фейнмана, и величины бегущего заряда $\alpha_3(\mu^2)$. Оба эти
параметра не фиксированы в теории, но зависимость $\alpha_3(\mu)$
такова, что $\Lambda_{\rm QCD}$ не зависит от численного значения
$\mu$. При $-q^2 = \Lambda_{\rm QCD}^2$ сильный заряд обращается в
бесконечность. Численно $\Lambda_{\rm QCD} \approx 300$ МэВ. Эта
величина определяет значения масс адронов, состоящих из легких $u$
и $d$ кварков.

С ростом $-q^2$ или уменьшением расстояний сильный заряд падает,
поэтому теория возмущений позволяет получить целый ряд
предсказаний для процессов с участием адронов, в которых
доминируют малые расстояния. Один из самых знаменитых примеров --
вычисление полного сечения аннигиляции $e^+ e^- \to$ адроны. Порог
реакции -- две массы $\pi^\pm$-мезонов, около 300 МэВ. С
увеличением энергии $e^+ e^-$-пары открываются новые каналы
реакции: $e^+ e^- \to 3\pi, 4\pi, \rho$, ... -- конечных состояний
все больше, и вычисление сечения становится все более сложной
задачей. Обходной путь -- вычислить сечение рождения кварковых пар
$u\bar u$, $d\bar d$ и $s\bar s$, которые на следующем этапе
реакции ``адронизируются''  -- превращаются в адроны, наблюдаемые
в конечном состоянии. Удобно рассматривать отношение сечения
рождения адронов к сечению рождения $\mu^+ \mu^-$-пар:
\begin{equation}
R \equiv \frac{\sigma_{e^+ e^-} \to {\rm hadrons}}{\sigma_{e^+
e^-} \to \mu^+ \mu^-} = 3(Q_u^2 + Q_d^2 +Q_s^2) = 3(\frac{4}{9} +
2 \cdot \frac{1}{9}) = 2 \;\; , \label{3.13}
\end{equation}
где множитель ``3'' происходит от трех цветов кварков, рождающихся
некогерентно в $e^+ e^-$-аннигиляции. Кварки имеют дробные
электрические заряды: $Q_u = Q_c = Q_t = +2/3$, $Q_d = Q_s = Q_b =
-1/3$. Предсказание независимости $R$ от полной энергии $e^+
e^-$-пары при $\sqrt s > 2$ ГэВ и число 2 хорошо подтверждаются
экспериментальными данными вплоть до $\sqrt s \approx 4$ ГэВ,
когда открывается возможность рождения содержащих $c$ ($\bar c$)
кварки очарованных адронов, и $R$ скачком возрастает до $\approx
3,3$. При дальнейшем увеличении до $\sqrt s \approx 10$ ГэВ $R$ не
меняется, а затем скачком возрастает до $\approx 3,6$ --
открывается возможность рождения ``прелестных'' адронов,
содержащих $b$ ($\bar b$) кварки. График зависимости
экспериментальных данных для $R$ от энергии можно найти в Обзоре
Физики Частиц.

Обсудим кварковый состав адронов. Составные модели адронов
восходят к модели Ферми начала 50-х годов, в которой недавно
открытые $\pi$-мезоны (переносчики сильных взаимодействий,
предсказанные Юкавой) считались составленными из пары нуклон -
антинуклон. Под нуклоном имеются ввиду протон или нейтрон, дублет
нуклонов ($p, n$) преобразуется по фундаментальному представлению
глобальной группы изоспина $SU(2)$, а находящиеся в присоединенном
триплетном представлении $SU(2)$ $\pi^+(p\bar n)$,
$\pi^0\left(\frac{\bar p p - n\bar n}{\sqrt 2}\right)$ и
$\pi^-(\bar p n)$ мезоны являются составными частицами. Открытие
странных частиц ($K$-мезоны, $\Lambda$, $\Sigma$, $\Xi$-гипероны)
потребовало расширения изоспиновой $SU(2)$ до флэйворной $SU(3)$
симметрии, в неприводимые представления которой комфортно
укладывались известные адроны (восьмиричный путь; Гелл-Манн и
Нееман). Наконец, в 1964 году Гелл-Манн и Цвейг ввели три кварка,
преобразующихся по фундаментальному представлению группы $SU(3)$
(Цвейг назвал их ``тузами'', но прижилось называние, данное
Гелл-Манном и почерпнутое им из романа Джойса). В современных
обозначениях это $u$ -- up, или верхний, $d$ -- down, или нижний,
и $s$ -- strange, или странный, кварки. Кварки имеют спин 1/2 и
могут находиться в одном их трех цветовых состояний, чему отвечает
индекс ``i'', скажем, $u^1$, $u^2$ и $u^3$, или $u^i$.

Проще всего инвариантные относительно цветовой группы $SU(3)_c$
``бесцветные'', или ``белые'', адроны построить из пары
кварк-антикварк: $\bar q_i q_i^\prime \equiv \bar q_1 q_1^\prime +
\bar q_2 q_2^\prime + \bar q_3 q_3^\prime$ (флэйворы $q$ и
$q^\prime$  в общем случае различны). Так устроены мезоны. В
основном состоянии орбитальный момент кварков равен нулю, и спин
мезона определяется полным спином пары кварк-антикварк. Аналогично
позитронию мы имеем состояния с $J^P = 0^-$, называемые
псевдоскалярными мезонами, и с $J^P = 1^-$ -- векторные мезоны.
Кварковый состав нонета псевдоскалярных мезонов: $\pi^+(u \bar
d)$, $\pi^0\left(\frac{u\bar u - d\bar d}{\sqrt 2}\right)$,
$\pi^-(d\bar u)$, $K^+(u\bar s)$, $K^0(d\bar s)$, $\bar K^0(s\bar
d)$, $K^-(s\bar u)$, $\eta_0\left(\frac{u\bar u + d\bar d - 2s\bar
s}{\sqrt 6}\right)$, $\eta_0^\prime\left(\frac{u\bar u + d\bar d +
s\bar s}{\sqrt 3}\right)$. Изосинглетное состояние $\eta_0$ и
$SU(3)$ синглет $\eta_0^\prime$ смешиваются, образуя физические
$\eta$- и $\eta^\prime$-мезоны. Масса $\pi$-мезонов близка к 140
МэВ, $K$-мезонов -- около 495 МэВ, $\eta$ -- 540 МэВ и, наконец,
$\eta^\prime$ весит 958 МэВ.

Нонет векторных мезонов: $\rho^+(u\bar d)$,
$\rho^0\left(\frac{u\bar u - d\bar d}{\sqrt 2}\right)$, $\rho^-(d
\bar u)$, $\omega\left(\frac{u\bar u + d\bar d}{\sqrt 2}\right)$,
$K^{*+}(u\bar s)$, $K^{*0}(d\bar s)$, $\bar K^{*0}(s\bar d)$,
$K^{*-}(s\bar u)$, $\phi(s\bar s)$. Векторные мезоны тяжелее
псевдоскалярных; масса $\rho$ близка к 770 МэВ, $\omega$ -- 780
МэВ, $K^*$ -- 890 МэВ, $\phi$ -- 1020 МэВ. Относительно большая
масса мезонов, в состав которых входит $s$-кварк, обусловлена
``тяжестью'' последнего. Псевдоскалярные мезоны распадаются по
слабым ($\pi \to \mu\nu$, $K\to \mu\nu$, $K\to\pi\pi$, $K\to
3\pi$) и электромагнитным ($\pi^0 \to 2\gamma$, $\eta \to
2\gamma$) взаимодействиям, что объясняет их сравнительно большое
время жизни. Поясним, почему $\eta$-мезон не распадается по
сильному взаимодействию. Распад $\eta\to 2\pi$ запрещен
$P$-четностью. Понять отсутствие сильного распада $\eta\to 3\pi$
помогает понятие $G$-четности, равной произведению зарядового
сопряжения $C$ на результат вращения на 180$^0$ вокруг оси ``y'' в
изопространстве. Изотопически и зарядовоинвариантное сильное
взаимодействие сохраняет $G$-четность, равную -1 для изотриплетных
$\pi$-мезонов и +1 для изосинглетных $\eta$- и
$\eta^\prime$-мезонов. $G$-четность системы трех $\pi$-мезонов
отрицательна, поэтому распады $\eta, \eta^\prime \to 3\pi$ не
могут идти по сильному взаимодействию (распады $\eta \to 3\pi$
идут по нарушающему изотопическую инвариантность электромагнитному
взаимодействию). Распад $\eta \to 4\pi$ запрещен кинематически,
так как масса системы $4\pi$ превышает массу $\eta$-мезона.

Векторные мезоны распадаются по сильному взаимодействию, что
объясняет их большую ширину. Распады $\rho\to 2\pi$ приводят к
ширине $\rho$-мезона $\Gamma_\rho \approx 150$ МэВ, что близко к
его массе. $G$-четность изоскалярного $\omega$-мезона
отрицательна, поэтому доминирует распад $\omega \to 3\pi$, и
$\Gamma_\omega \approx 10$ МэВ. Распады $K^* \to K\pi$ приводят к
$\Gamma_{K^*} \approx 50$ МэВ, и, наконец, распад $\phi \to K\bar
K$ дает $\Gamma_\phi \approx 4$ МэВ (малая ширина обусловлена
маленьким фазовым объемом рождающихся $K$-мезонов, $v_K =
\sqrt{1-(2m_K/m_\phi)^2} \approx 0.25$).

Барионы состоят из антисимметричной по цвету комбинации трех
кварков. Поэтому в силу принципа Паули произведение координатной,
спиновой и флэйворной волновых функций кварков  в барионах должно
быть симметрично относительно перестановки кварковых полей.
Орбитальные моменты кварков в основном состоянии равны нулю
(пространственная волновая функция тем самым симметрична).
Симметричная спиновая волновая функция отвечает спину бариона 3/2
и требует симметричной флэйворной волновой функции. Спину бариона
1/2 отвечает смешанная симметрия спиновой волновой функции.
Произведение трех $SU(3)$ флэйворных волновых функций кварков дает
27 компонентов, из которых 10 принадлежат симметричному
представлению (декуплет), 1 -- антисимметричному и две восьмерки
отвечают смешанному представлению (октеты).

Приведем кварковый состав октета барионов:

\bigskip

\begin{center}
\begin{tabular}{ccccc}
& $p(uud)$ & & $n(udd)$ & \\
&&&& \\
&&&& \\
&&$\Lambda([ud]s)$ & & \\
$\Sigma^+(uus)$ & & $\Sigma^0(\{ud\}s)$ & & $\Sigma^-(dds)$  \\
&&&& \\
&&&& \\
 & $\Xi^0(uss)$ & & $\Xi^-(dss)$ &
\end{tabular}
\end{center}

\bigskip

Здесь симметричная комбинация $u$- и $d$-кварков обозначена
фигурными скобками (изоспин $\Sigma$-гиперонов равен единице),
антисимметричная -- квадратными (изоспин $\Lambda$-гиперона равен
нулю). Массы протона и нейтрона близки к 940 МэВ, $m(\Lambda)
\approx 1115$ МэВ, $m(\Sigma) \approx 1190$ МэВ, $m(\Xi) \approx
1320$ МэВ. Более тяжелые гипероны распадаются на более легкие по
меняющему флэйвор кварков слабому взаимодействию, протон же, как
самый легкий барион, стабилен (распады типа $p\to e^+ \pi^0$
запрещены законом сохранения барионного числа).

Декуплет гиперонов со спином 3/2 приведен вместе с кварковым
составом и массами частиц:

\bigskip

\begin{center}

\begin{tabular}{cccccccc}
$\Delta^{++}(uuu)$ & & $\Delta^+(uud)$ & & $\Delta^0(udd)$ & &
$\Delta^-(ddd)$ & 1230 МэВ \\
&&&&&&& \\
&&&&&&& \\
& $\Sigma^{+*}(uus)$ & & $\Sigma^{0*}(uds)$ & & $\Sigma^{-*}(dds)$
&& 1385 МэВ \\
&&&&&&& \\
&&&&&&& \\
&& $\Xi^{0*}(uss)$ & & $\Xi^{-*}(dss)$ & && 1530 МэВ \\
&&&&&&& \\
&&&&&&& \\
&&& $\Omega^-(sss)$ &&& &1670 МэВ
\end{tabular}

\end{center}

\bigskip

Существование $\Omega^-$-гиперона было предсказано Гелл-Манном,
исходя из $SU(3)$ симметрии.

Члены декуплета распадаются на гипероны из октета с испусканием
$\pi$-мезона; распады идут за счет сильного взаимодействия.
Единственное исключение -- $\Omega^-$ -гиперон, распадающийся по
слабому взаимодействию. Время жизни $\Sigma$, $\Lambda$, $\Xi$ и
$\Omega$ гиперонов -- порядка $10^{-10}$ секунды, а ширины
$\Delta$, $\Sigma^*$ и $\Xi^*$ варьируются от 100 МэВ ($\Delta$)
до 10 МэВ ($\Xi^*$). Большое время жизни нейтрона ($\approx$ 900
секунд) объясняется малым энерговыделением в распаде $n\to pe^-
\bar\nu_e$. Электрические заряды кварков однозначно следуют из
зарядов $\Delta^{++} (Q_u = +2/3)$, $\Delta^- (Q_d = -1/3)$ и
$\Omega^- (Q_s = -1/3)$. При этом мы полагаем, что заряд кварка не
зависит от его цвета.

Массы кварков определяются косвенным образом. Начнем с легких
$u$-, $d$- и $s$-кварков. Доминирующий распад заряженного
$\pi^-$-мезона $\pi^- \to \mu^-\nu_\mu$ происходит через
аннигиляцию $\pi^-$-мезона аксиальным кварковым током, описываемую
следующим матричным элементом
\begin{equation}
\langle 0 | \bar u \gamma_\alpha \gamma_5 d | \pi^- \rangle =
f_\pi p_\alpha \;\; , \label{3.14}
\end{equation}
где $p_\alpha$ -- 4-импульс $\pi$-мезона. Домножая обе части
(\ref{3.14}) на $p_\alpha$ и пользуясь уравнением Дирака для
кварковых полей, получим:
\begin{equation}
(m_d + m_u) \langle 0 | \bar u \gamma_5 d | \pi^- \rangle = f_\pi
m_\pi^2 \;\; . \label{3.15}
\end{equation}
Из аналогичного рассмотрения распада $K^-$-мезона $K^-
\to\mu\nu_\mu$ получим:
\begin{equation}
(m_s + m_u) \langle 0 | \bar u \gamma_5 s | K^-\rangle = f_K m_K^2
\;\; . \label{3.16}
\end{equation}
В силу флэйворной $SU(3)$ симметрии $f_\pi \approx f_K$, и стоящие
в (\ref{3.15}) и (\ref{3.16}) матричные элементы также примерно
равны, что приводит к следующему результату:
\begin{equation}
\frac{m_s + m_u}{m_d + m_u} = \frac{m_K^2}{m_\pi ^2} \approx 13
\;\; . \label{3.17}
\end{equation}
Масса странного кварка может быть определена из разности масс
членов декуплета гиперонов:
\begin{equation}
m_s \approx m_{\Sigma^*} - m_\Delta \approx m_{\Xi^*} -
m_{\Sigma^*} \approx m_\Omega - m_{\Xi^*} \approx 150 \; \mbox
{\rm МэВ} \;\; . \label{3.18}
\end{equation}
Если массы $u$- и $d$-кварков были бы равны, то из (\ref{3.17}) и
(\ref{3.18}) мы бы нашли, что $m_u = m_d = 6$ МэВ. Анализ
разностей масс членов изомультиплетов позволяет определить
разность масс $u$- и $d$-кварков. В результате было получено:
\begin{equation}
m_u \approx 3 \; \mbox{\rm МэВ} \; , \;\; m_d \approx 7 \;
\mbox{\rm МэВ} \; , \;\; m_s \approx 150 \; \mbox{\rm МэВ} \;\; .
\label{3.19}
\end{equation}

Естественно возникает вопрос: почему в силу $SU(3)$-симметрии
$f_\pi \approx f_K$, в то время как $m_\pi^2$ и $m_K^2$ столь
сильно отличаются? Тут дело в специальной роли псевдоскалярных
мезонов, массы которых зануляются в пределе безмассовых кварков.
Именно поэтому легкость $\pi$-мезона ($m_\pi^2 \approx 0.02$
ГэВ$^2$) свидетельствует о малости масс $u$- и $d$-кварков
(имеется ввиду малость в масштабе $\Lambda_{\mbox{\rm КХД}}
\approx 300$ МэВ).

Перейдем к тяжелым $c$-, $b$- и $t$-кваркам, массы которых велики
по сравнению с $\Lambda_{\mbox{\rm КХД}}$. Масса содержащих $c$- и
$b$-кварки адронов в основном определяется массами тяжелых
кварков. Состоящий из $c\bar c$-кварков $J/\psi$ мезон весит 3.1
ГэВ, поэтому масса $c$-кварка примерно равна 1.5 ГэВ. Состоящий из
$b\bar b$-кварков $\Upsilon$-мезон весит 9.4 ГэВ; масса $b$-кварка
близка к 4.5 ГэВ. $t$-кварк весит около 174 ГэВ и за ядерное время
распадается, поэтому адронных состояний он образовывать не
успевает. Его масса определяется из кинематики его распада (сумма
4-импульсов рождающихся частиц в квадрате равна квадрату массы
$t$-кварка).

В заключение сделаем следующее замечание. Говорить о том, что
адрон составлен из каких-то определенных легких кварков не совсем
корректно. Кварки испускают глюоны, которые, в свою очередь, могут
рождать кварк-антикварковые пары. Так как константа сильных
взаимодействий $\alpha_s$ велика на адронном масштабе, такими
процессами нельзя пренебрегать. Поэтому была выработана подходящая
терминология: кварки блоковые (из которых состоят адроны) и кварки
токовые (входящие в лагранжиан КХД). Блоковые $u$- и $d$-кварки
весят примерно по 300 МэВ (их утроенная масса близка к массам
нуклонов). Мы же определяли величины масс токовых кварков.

\newpage

\begin{center}

{\bf Лекция 4 \\
$\mbox{$\mathbf{SU(2)_L \times U(1)}$}$ -- электрослабая теория.}

\end{center}

\setcounter{equation}{0} \def\theequation{4.\arabic{equation}}

\bigskip

Первую теорию слабых взаимодействий, объясняющую $\beta$-распады
ядер четырехфермионным переходом $n\to p e^- \bar\nu_e$, предложил
в 1934 году Энрико Ферми. Электрон и нейтрино при этом образуются
в ходе распада аналогично образованию фотона при распаде
возбужденного атома.

Описывающий $\beta$-распад нейтрона лагранжиан имеет следующий
вид:
\begin{equation}
{\cal L} = \frac{G_F}{\sqrt 2} \bar p \gamma_\alpha (1+ g_A
\gamma_5) n \bar e \gamma_\alpha (1+\gamma_5) \nu_e \;\; ,
\label{4.1}
\end{equation}
где константа взаимодействия $G_F \approx 10^{-5}/m^2_p$, и ее
малость объясняет слабость взаимодействия.

Токи $\bar e \gamma_\alpha \nu_e$ и $\bar p \gamma_\alpha n$
являются векторными, а $\bar e \gamma_\alpha \gamma_5 \nu_e$ и
$\bar p \gamma_\alpha \gamma_5 n$, где
\begin{equation}
\gamma_5 \equiv -i \gamma_0 \gamma_1 \gamma_2 \gamma_3 = \left(
\begin{array}{rr}
0 & -1 \\
-1 & 0
\end{array}
\right) \;\; \mbox{--}  \label{4.2}
\end{equation}
-- аксиальными. Поэтому в (\ref{4.1}) наряду с $P$-четными членами
$V \cdot V$ и $A \cdot A$ входит $P$-нечетная комбинация $V \cdot
A$ -- слабое взаимодействие нарушает пространственную четность.
Нарушающее $P$-четность четырехфермионное взаимодействие было
установлено во второй половине 50-х годов; Ферми в своей теории
использовал произведение векторных токов по аналогии с квантовой
электродинамикой, отмечая, тем не менее, возможность других
вариантов.

Выясним физический смысл проектора $P_L = (1+\gamma_5)/2$.
Записывая биспинор через двухкомпонентные спиноры, мы видим, что
$P_L$ выделяет суперпозицию $\varphi - \chi$. Из уравнения Дирака
(\ref{2.13}) получим:
\begin{equation}
\varphi - \chi = \left( 1-\frac{\bar\sigma \bar
p}{E+m}\right)\varphi \;\; . \label{4.3}
\end{equation}

В случае ультрарелятивистской частицы $E \gg m$ (нейтрино
удовлетворяет этому требованию во всех практически интересных
случаях), предполагая, что частица движется вдоль оси $z$, мы
убеждаемся, что ее спин ориентирован против импульса:
\begin{equation}
(1-\bar\sigma \bar n) \left(
\begin{array}{c}
\varphi_+ \\
\varphi_-
\end{array}
\right) = \left(
\begin{array}{cc}
0 & 0 \\
0 & 2
\end{array}
\right)\left(
\begin{array}{c}
\varphi_+ \\
\varphi_-
\end{array}
\right) = 2 \left(
\begin{array}{c}
0 \\
\varphi_-
\end{array}
\right) \;\; . \label{4.4}
\end{equation}
Таким образом, проектор $P_L$ выделяет поляризованное против
импульса нейтрино, рождающееся в слабых взаимодействиях. Очевидно
нарушение $P$-четности: при отражении координат $\bar x \to -\bar
x$ изменяет направление импульс нейтрино, а спин, являясь
аксиальным вектором, направления не меняет. Поэтому в
``зазеркальном мире'' в слабых распадах нейтронов рождаются
поляризованные вдоль импульса нейтрино.

Биспинор $\psi_L \equiv P_L \psi$ называют левым, а $\psi_R \equiv
P_R \psi$ ($P_R = (1-\gamma_5)/2$) -- правым. В
ультрарелятивистском случае $\psi_L$ описывает поляризованную
против импульса частицу, а $\psi_R$ -- частицу, поляризованную
вдоль импульса.

С точки зрения современной теории распад нейтрона происходит за
счет изображенного на рис. 4.1 распада $d$-кварка $d\to u e^-
\bar\nu_e$, обусловленного обменом промежуточным заряженным
$W$-бозоном.

\begin{center}
\bigskip
\includegraphics[width=.4\textwidth]{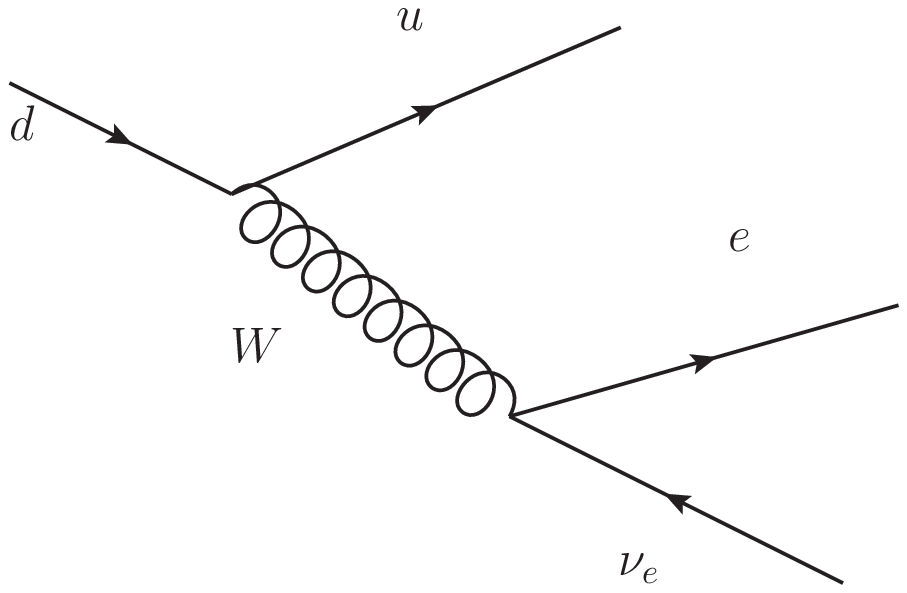}

\vspace{5mm}

{\it Рис. 4.1. Распад $d$-кварка}

\end{center}

\bigskip

Описываемая диаграммой рис. 4.1 амплитуда сводится к
четырехфермионной и дает микроскопическое объяснение происхождения
фермиевской константы:
\begin{equation}
G_F \sim \frac{g^2}{M_W^2} \approx \frac{(0.3)^2}{(80 \mbox{\rm
ГэВ})^2} \;\; . \label{4.5}
\end{equation}

До сих пор мы говорили о распадах частиц, которые обусловлены
обменом заряженным $W$-бозоном. Создаваемые на ускорителях пучки
мюонных нейтрино $\nu_\mu$ позволили изучить реакцию неупругого
рассеяния $\nu_\mu N \to \mu^- X$, где $N$ -- ядро мишени, а $X$
-- состояние, содержащее большое количество сильно
взаимодействующих частиц (адронов). В 1973 году в ЦЕРНе были
обнаружены аналогичные события без рождения мюона. Эти события
получили название ``нейтральных токов'' -- они обусловлены обменом
нейтральным $Z$-бозоном, смотри рис. 4.2.

\begin{center}
\bigskip
\includegraphics[width=1.\textwidth]{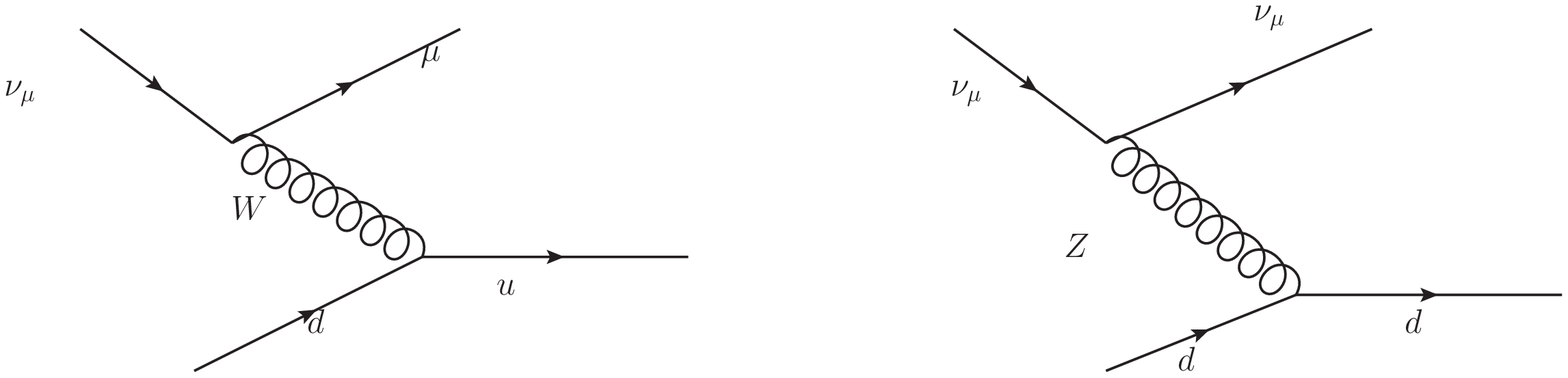}

\vspace{5mm}

{\it Рис. 4.2. Диаграммы, приводящие к неупругому рассеянию за
счет заряженного ($W$) и нейтрального ($Z$) токов}

\end{center}

\bigskip

Итак, слабые взаимодействия обусловлены обменами тяжелыми
промежуточными $W^\pm$- и $Z$-бозонами. Тем самым теория (как и
предвидел Ферми) близка к квантовой электродинамике, отличаясь от
нее в двух пунктах: в отличие от фотона, $W$ и $Z$ массивны, и
теория неабелева -- имеется три промежуточных векторных бозона:
$W^+$, его античастица $W^-$ и нейтральный $Z$. Однако придать
самосогласованным образом массы промежуточным бозонам оказалось
непросто.

Начнем с $U(1)$-теории и добавим в лагранжиан массовый член:
\begin{equation}
{\cal L} = -\frac{1}{4} F_{\mu\nu}^2 + \frac{1}{2} M^2 A_\mu^2
\;\; . \label{4.6}
\end{equation}
Уравнение Лагранжа
\begin{equation}
\frac{\partial}{\partial x_\mu} \frac{\delta {\cal L}}{\delta
\partial_\mu A_\nu} = \frac{\delta {\cal L}}{\delta A_\nu}
\label{4.7}
\end{equation}
приводит к следующему уравнению на поле $A_\mu$:
\begin{equation}
\square A_\mu - \partial_\mu \partial_\nu A_\nu + M^2 A_\mu = 0
\;\; . \label{4.8}
\end{equation}
Уравнение на функцию Грина в импульсном представлении имеет
следующий вид:
\begin{equation}
\left[(k^2 - M^2) g_{\mu\nu} - k_\mu k_\nu\right] G_{\nu\rho} =
\delta_{\mu\rho} \;\; , \label{4.9}
\end{equation}
решая которое, для функции Грина (или пропагатора) векторной
частицы получим:
\begin{equation}
G_{\mu\nu} = \frac{g_{\mu\nu} - \frac{k_\mu k_\nu}{M^2}}{k^2 -
M^2} \;\; . \label{4.10}
\end{equation}

Импульс, текущий по пропагатору $W$ при слабых распадах частиц
$|k| \ll M$, и пропагатор ведет себя как $1/M^2_W$, что объясняет
оценку (\ref{4.5}). Однако в пределе больших импульсов $|k| \gg M$
пропагатор не падает с ростом импульса, что делает теорию с
массивным векторным бозоном неперенормируемой. Последнее означает,
что петлевые поправки к древесным амплитудам расходятся
неконтролируемым образом, и, тем самым, успешное описание распадов
частиц древесными амплитудами не имеет теоретического обоснования.
Вычисление амплитуд в неперенормируемой теории невозможно. Корень
зла -- массовый член в лагранжиане (\ref{4.6}). Возникает задача
придания массы векторным бозонам без введения массового члена в
лагранжиан. Эта задача была решена практически одновременно в
нескольких теоретических работах в 1964 году, когда было
обнаружено явление, позже названное эффектом Хиггса по имени
автора одной из этих работ. Аналогичное физические явление имеет
место в теории сверхпроводимости Гинзбурга--Ландау (1950). В этой
теории выталкивание магнитного поля из сверхпроводника (эффект
Мейснера) происходит за счет смешивания фотона с безмассовым
скалярным полем (параметром порядка) $\varphi$, в результате чего
фотон становится массивным и магнитное поле экспоненциально
затухает вглубь сверхпроводника.

Безмассовое скалярное поле -- очень неестественный объект:
радиационные поправки немедленно сделают такое поле массивным.
Поэтому изложение перенормируемой теории массивных векторных полей
(а это теория слабых взаимодействий) следует начать с эффекта
Голдстоуна, позволяющего получать безмассовые скалярные частицы
естественным образом.

Рассмотрим теорию комплексного скалярного поля, описываемую
следующим лагранжианом:
\begin{equation}
{\cal L} = |\partial_\mu \phi|^2 - \lambda^2[|\phi|^2 -\eta^2]^2
\;\; . \label{4.11}
\end{equation}
Эта теория инвариантна относительно глобальных
$U(1)$-преобразований: $\phi(x) = e^{i\alpha}\phi^\prime(x)$. Если
имеющий размерность квадрата массы параметр $\eta^2$ отрицателен,
то мы имеем дело с массивным комплексным полем $\phi$, лагранжиан
для которого без члена $\sim|\phi|^4$ нам уже встречался, см.
формулу (\ref{2.5}). Гораздо более интересен случай $\eta^2 > 0$,
когда минимум потенциала достигается на окружности $|\phi| =
\eta$, а сам потенциал $V(\phi) = \lambda^2[(\phi)^2 - \eta^2]^2$
является фигурой вращения и напоминает мексиканскую шляпу
``сомбреро'', или донышко бутылки, смотри рис. 4.3.

\begin{center}
\bigskip
\includegraphics[width=.6\textwidth]{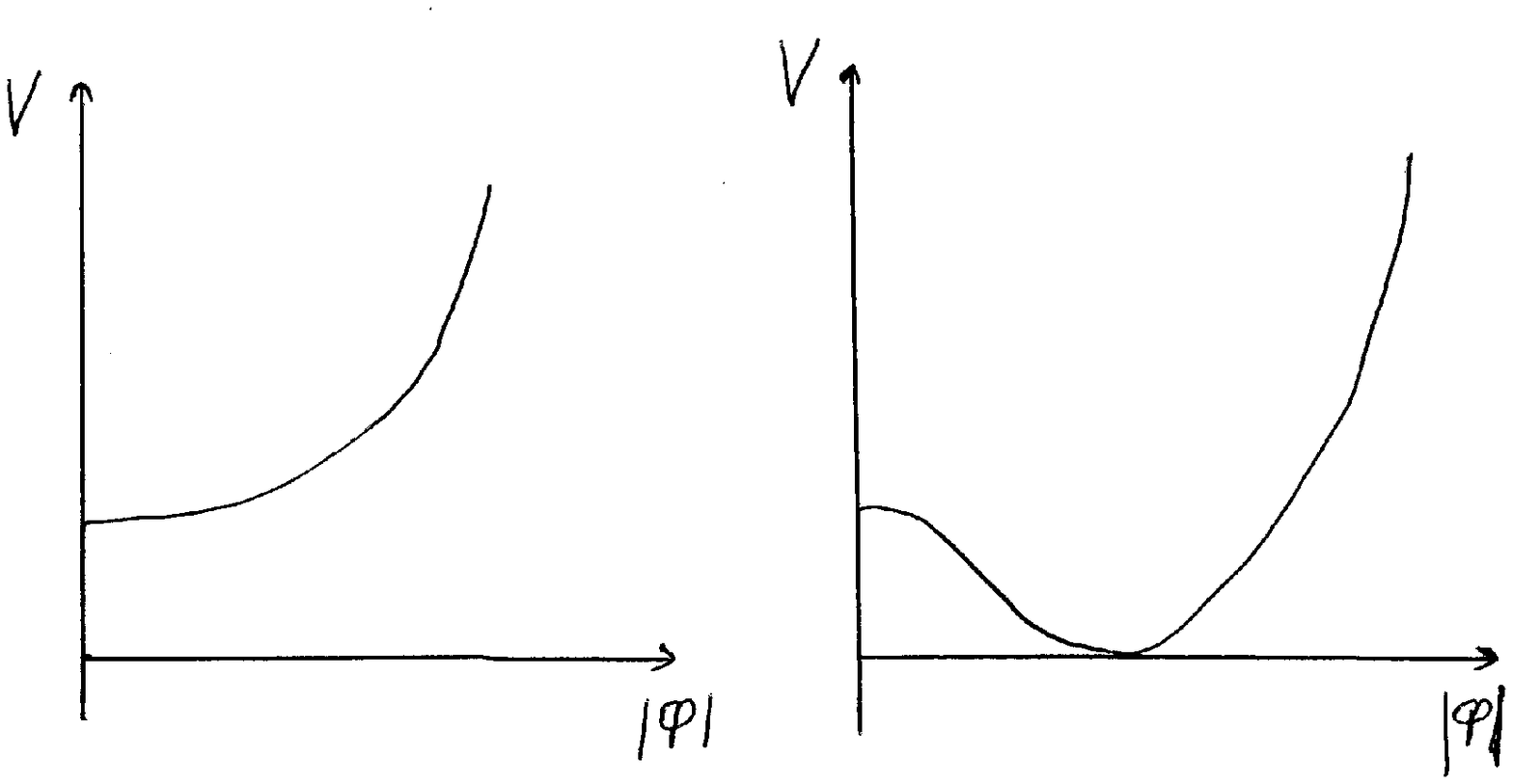}

\vspace{5mm}

{\it Рис. 4.3. Зависимость $V(\phi)$}

\end{center}

\bigskip

Выбирая в качестве вакуума состояние $<\phi> = \eta$ и раскладывая
поле $\phi$ в окрестности вакуума
\begin{equation}
\phi(x) = \eta + \rho(x) + i\varphi(x) \;\; , \label{4.12}
\end{equation}
получим:
\begin{equation}
{\cal L} = (\partial_\mu \rho)^2 - 4\eta^2 \lambda^2 \rho^2 +
(\partial_\mu \varphi)^2 - 4\eta \lambda^2 \rho \varphi^2 -
\lambda^2 \varphi^4 \;\; . \label{4.13}
\end{equation}
Комплексное поле $\phi$ распалось на два вещественных: $\rho$ с
массой $2\lambda\eta$ и безмассовое $\varphi$. Произошло явление,
не совсем удачно называемое спонтанным нарушением симметрии:
вакуум теории не инвариантен относительно $U(1)$-преобразования:
$\eta = e^{i\alpha}\eta^\prime$. Тем не менее, $U(1)$-симметрия
осталась: именно из-за нее поле $\varphi$ не имеет массы. Не
возникает масса у $\varphi$ и при учете радиационных поправок. С
таким явление в квантовой теории поля в конце 50-х годов
встретился Джефри Голдстоун; безмассовая частица, отвечающая полю
$\varphi$, получила название голдстоуновского бозона.

Следующий шаг -- сделать $U(1)$-симметрию локальной, ``удлинив''
производную в кинетическом члене (\ref{4.11}) и добавив лагранжиан
абелева векторного поля:
\begin{equation}
{\cal L} = |D_\mu \phi|^2 - \frac{1}{4} F_{\mu\nu}^2 -
\lambda^2[|\phi|^2 - \eta^2]^2 \;\; . \label{4.14}
\end{equation}
Подставляя разложение около минимума потенциала скалярного поля
(\ref{4.12}), увидим, что голдстоуновский бозон из-за члена $e\eta
\partial_\mu \varphi A_\mu$ смешался с векторным бозоном, а у того
возник массовый член $e^2 \eta^2 A_\mu^2$. Из двух поляризаций
безмассового векторного поля $A_\mu$ и двух степеней свободы
комплексного поля $\phi$ получилось массивное векторное поле,
имеющее три поляризации, и вещественный массивный скаляр $\rho$.
Голдстоуновское поле $\varphi$ поглощено безмассовым векторным
полем и служит его третьей компонентой поляризации. За счет этого
поглощения $A_\mu$ стало массивным. Численное значение вакуумного
среднего $\eta$ определяется величиной константы слабого
взаимодействия: $G_F \sim e^2/(e^2 \eta^2) = 1/\eta^2$, $\eta =
246$ ГэВ.

Наконец все готово для изложения электрослабой теории
Глэшоу--Вайнберга--Салама, основанной на локальной симметрии
$SU(2) \times U(1)$. В теории имеется триплет векторных полей
$A_\mu^i$, $i=1,2,3$ и абелево поле $B_\mu$, калибрующие,
соответственно, $SU(2)$- и $U(1)$-симметрии. Ковариантная
производная имеет следующий вид:
\begin{equation}
D_\mu = \partial_\mu - ig A_\mu^i T^i - ig^\prime \frac{Y}{2}B_\mu
\;\; , \label{4.15}
\end{equation}
где $g$ и $g^\prime$ -- заряды, а $Y$ -- гиперзаряд, подбираемый
таким образом, чтобы кварки и лептоны имели наблюдаемые на
эксперименте электрические заряды. Лагранжиан бозонного сектора
имеет следующий вид:
\begin{equation}
{\cal L}_1 = -\frac{1}{4}B_{\mu\nu}^2 - \frac{1}{2}{\rm Tr}
A_{\mu\nu}^2 + |D_\mu H|^2 - \lambda^2 [|H|^2 - \eta^2]^2 \;\; ,
\label{4.16}
\end{equation}
где дублет хиггсов $H = (H^+, H^0)$ приобретает в минимуме
потенциала вакуумное среднее $<H> = (0, \eta)$. При этом
калибровочная $SU(2) \times U(1)$-симметрия нарушается до абелевой
$U(1)$-группы, описывающей квантовую электродинамику. Четыре
векторных бозона $A_\mu^i$ и $B_\mu$ образуют заряженный бозон
$W_\mu^\pm = (A_\mu^1 \pm iA_\mu^2)/\sqrt 2$ и нейтральные бозоны:
массивный $Z_\mu$ и безмассовый фотон $A_\mu$. Два последних
являются линейными суперпозициями $A_\mu^3$ и $B_\mu$. Параметры
модели $\eta$, $g$ и $g^\prime$ фиксируются значениями $G_F$, $e$
и $M_Z$, известными с очень хорошей точностью. $W$- и $Z$-бозоны
были обнаружены в ЦЕРНе в 1983 году на протон-антипротонном
коллайдере SPS. Последний неизвестный параметр $\lambda$
определяет массу бозона Хиггса. Обнаружение в 2012 году бозона
Хиггса с массой 126 ГэВ на протон-протонном коллайдере LHC в ЦЕРНе
зафиксировало численное значение $\lambda$. Это случилось через 45
лет после публикации работы Вайнберга, содержавшей электрослабую
$SU(2) \times U(1)$-модель с хиггсовским механизмом генерации масс
промежуточных векторных бозонов (работа Глэшоу опубликована в 1961
году, Салама -- в 1969).

Перейдем к фермионному сектору. Левые кварки и лептоны входят в
$SU(2)$-дублеты, правые являются синглетами:
\begin{equation}
\left(
\begin{array}{c} u \\
d
\end{array}
\right)_L \; , \;\; \left(
\begin{array}{c} c \\
s
\end{array}
\right)_L \; , \;\; \left(
\begin{array}{c} t \\
b
\end{array}
\right)_L \; ;
\begin{array}{ccc}
u_R , & c_R , & t_R \\
d_R , & s_R , & b_R
\end{array}
\label{4.17}
\end{equation}

\begin{equation}
\left(
\begin{array}{c} \nu_1 \\
e
\end{array}
\right)_L \; , \;\; \left(
\begin{array}{c} \nu_2 \\
\mu
\end{array}
\right)_L \; , \;\; \left(
\begin{array}{c} \nu_3 \\
\tau
\end{array}
\right)_L \; ;
\begin{array}{ccc}
e_R , & \mu_R , & \tau_R \\
N_{1_R} , & N_{2_R} , & N_{3_R}
\end{array}
\label{4.18}
\end{equation}

На всякий случай мы ввели в состав частиц правые нейтрино $N_i$,
хотя, возможно, таких частиц и нет в природе. Кинетические члены
первого, второго и третьего семейства одинаковы; выпишем их для
первого семейства:
\begin{equation}
{\cal L}_2 = \bar L \hat D L + \bar Q \hat D Q + \bar e_R \hat D
e_R + \bar u_R \hat D u_R + \bar d_R \hat D d_R + \hat N_{1_R}
\hat D N_{1_R} \;\; , \label{4.19}
\end{equation}
где буквами $L$ и $Q$ обозначены лептонный и кварковый дублет,
соответственно. Структура лагранжиана такова, что с $W$-бозонами
взаимодействуют только левые фермионы, а с фотоном и $Z$-бозоном
-- и левые, и правые, причем с фотоном левые и правые
взаимодействуют с одной и той же константой связи, и поэтому
электромагнитный ток чисто векторный.

Осталась неразобранной последняя часть лагранжиана, дающая массы
кваркам и лептонам. Это юкавское взаимодействие хиггсовского
дублета с фермионами. Если бы Стандартная Модель содержала одно
кварк-лептонное семейство, то мы имели бы
\begin{equation}
\tilde{\cal L}_3 = f_l \bar L H e_R + f_d \bar Q H d_R + f_u \bar
Q \tilde H u_R \;\; , \label{4.20}
\end{equation}
где $f_e = m_e/\eta$, $f_d = m_d/\eta$, $f_u = m_u/\eta$, а
$\tilde H_a \equiv \varepsilon_{ab} H_b^*$ и $\varepsilon_{ab} =
\left(\begin{array}{cc} 0& 1 \\
-1 & 0
\end{array}\right)$. При подстановке в (\ref{4.20}) $H = (0, \eta
+ H(x))$ мы получаем массовые члены фермионов, а также их
взаимодействия с бозоном Хиггса. Последние пропорциональны массам
фермионов: чем тяжелее фермион, тем больше константа его
взаимодействия с хиггсовским бозоном. Обсуждение массы нейтрино мы
отложим до следующей лекции.

Может возникнуть следующий вопрос: $SU(2)$ -- неабелева группа; не
возникает ли в ней конфайнмента, ведущего к ``запиранию''
промежуточных бозонов и кварков с лептонами аналогично тому, как
это происходит с кварками и глюонами в $SU(3)_c$-теории сильных
взаимодействий. Ответ -- нет, и это видно из формулы для
пропагатора $W$-бозона с учетом рад. поправок, схематично имеющей
следующий вид:
\begin{equation}
G_W(q^2) = \frac{1}{q^2 - M_W^2 + \alpha_2(M_W^2) q^2
\ln\left(\frac{-q^2}{M_W^2}\right)} \;\; . \label{4.21}
\end{equation}
При $-q^2 \gg M_W^2$ массой $W$-бозона в знаменателе можно
пренебречь, и с уменьшением $-q^2$ пропагатор растет, как и
положено в неабелевой теории. Однако когда $-q^2$ становится
порядка $M_W^2$ и уменьшается дальше, в знаменателе доминирует
член $M_W^2$ -- рост пропагатора прекращается, и в рассматриваемом
случае он ``замерзает'' при численно малом значении: в хиггсовской
фазе мы имеем дело с последовательной неабелевой теорией с малой
константой связи.

Обсудив почти все ингредиенты электрослабой теории, перейдем к
анализу физических процессов и начнем с открытия бозона Хиггса на
LHC.

Сталкивающиеся в LHC протоны содержат легкие кварки, связь которых
с бозоном Хиггса чрезвычайно мала: $m_q /\eta \sim 10^{-5}$.
Поэтому основная часть хиггсовских бозонов рождается за счет
механизма, показанного на рис. 4.4: излучаемые легкими кварками
глюоны рождают виртуальные (находящиеся вне массовой поверхности)
$t$-кварки, которые в силу своей очень большой массы сильно
связаны с $H$.

\newpage

\begin{center}

\includegraphics[width=.4\textwidth]{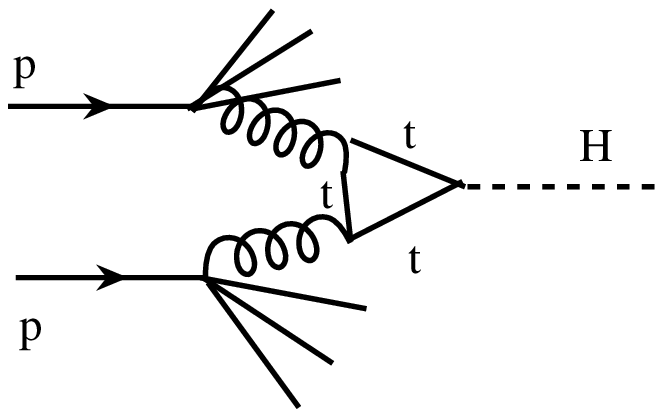}

\vspace{5mm}

{\it Рис. 4.4. Рождение хиггсовского бозона на LHC}

\end{center}

\bigskip

Но родить новую частицу -- это лишь полдела. Необходимо ее
задетектировать. Ширина $H$ близка к 5 МэВ, что отвечает времени
жизни $\sim 10^{-22}$ секунды. Поэтому $H$ распадается практически
в той же точке, где и рождается, и его детектирование происходит
по продуктам распада. В настоящее время измерены произведения
сечения рождения $H$ на относительные вероятности его распадов по
пяти каналам: $H \to WW^*$, $ZZ^*$, $b\bar b$, $\tau^+ \tau^-$ и
$\gamma\gamma$ (см. рис. 4.5 - 4.8). Результаты совпадают с
предсказаниями теории, хотя точность пока невелика, порядка 30\%.
Точность экспериментальных данных заметно повысится после набора
данных на LHC в 2015 - 2017 годах.

\begin{center}
\bigskip
\includegraphics[width=.4\textwidth]{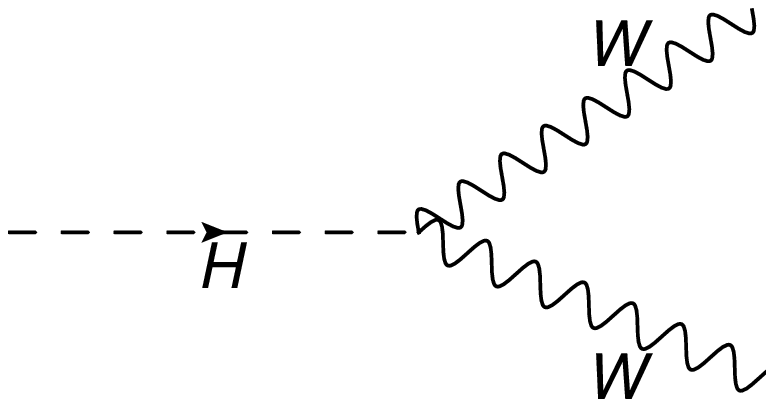}

\vspace{5mm}

{\it Рис. 4.5. Распад $H\to WW^*$}

\end{center}

\newpage

\begin{center}

\bigskip

\includegraphics[width=.4\textwidth]{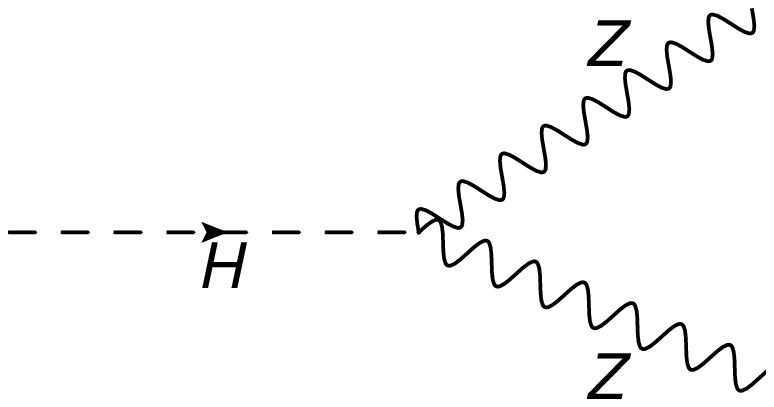}

\vspace{5mm}

{\it Рис. 4.6. Распад $H\to ZZ^*$}

\end{center}

\bigskip

\begin{center}

\bigskip

\includegraphics[width=.4\textwidth]{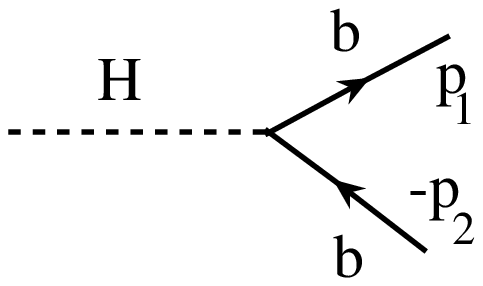}

\vspace{5mm}

{\it Рис. 4.7. Распад $H\to b\bar b$. Такая же диаграмма с заменой
$b\to \tau$ описывает распад $H\to\tau\tau$}

\end{center}

\bigskip

\begin{center}

\bigskip

\includegraphics[width=.8\textwidth]{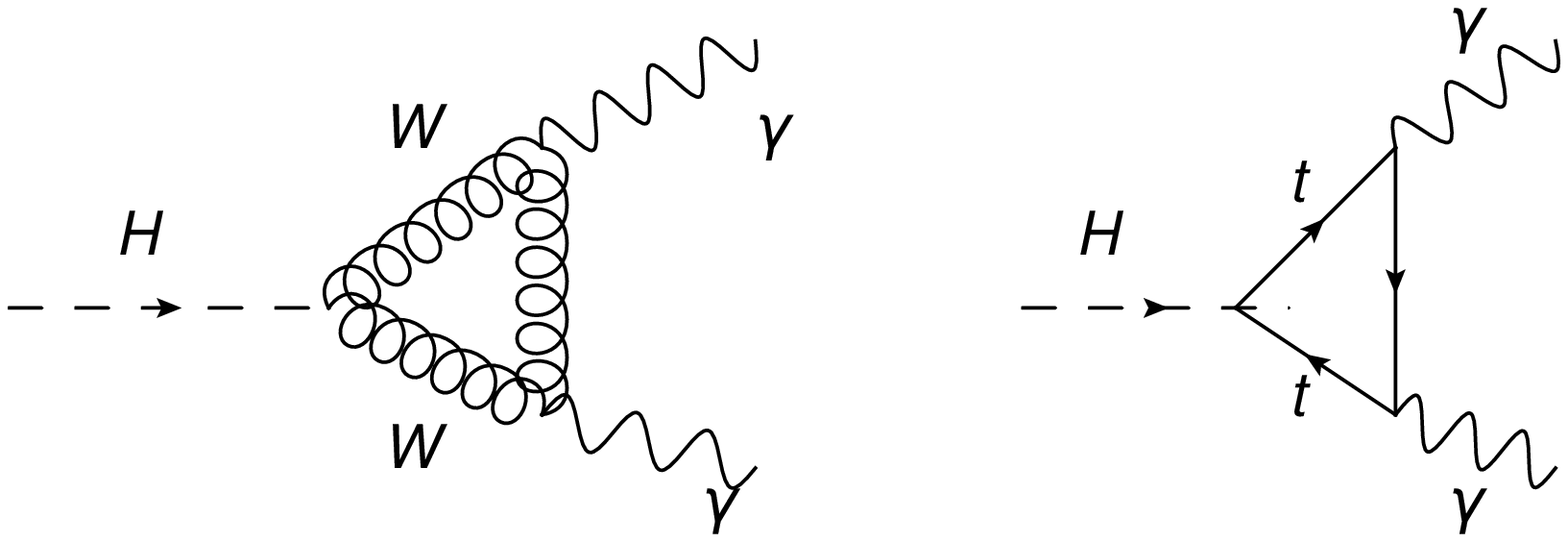}

\vspace{5mm}

{\it Рис. 4.8. Распад $H\to 2\gamma$}

\end{center}

\bigskip

До 1956 года считалось, что наряду с непрерывной группой
преобразований Лоренца квантовая теория поля инвариантна
относительно дискретных преобразований отражения пространственных
координат $\bar x \to -\bar x$ ($P$), времени $t\to -t$ ($T$) и
замены частиц на античастицы ($C$). Это было так для наиболее
хорошо изученной теории -- квантовой электродинамики. Сейчас мы
знаем, что ненарушенной является только симметрия CPT --
произведение трех дискретных симметрий: симметрии P, C, CP, T
нарушаются в слабом взаимодействии.

Все началось с так называемой $\theta - \tau$ проблемы: были
обнаружены распады $\theta^+ \to (2\pi)^+$ и $\tau^+ \to
(3\pi)^+$, которые приписывали разным частицам $\theta$ и $\tau$.
Это было вызвано тем, что P-четность системы $2\pi$ в $s$-волне
положительна, а системы $3\pi$ -- отрицательна. Инвариантные массы
$\theta$ и $\tau$ оказались равными; более того, они имели
одинаковое время жизни. Ли и Янг предположили, что $\theta$ и
$\tau$ -- это одна и та же частица, в распадах которой P-четность
нарушается. Сейчас эта частица называется $K^+$-мезоном. Нарушение
P-четности в слабых распадах вскоре было подтверждено в опытах по
распадам ядер, где была найдена корреляция спина распадающегося
ядра и импульса вылетающего электрона вида $\vec s \vec p$,
которая явно нарушает P-четность, так как импульс $\vec p$ --
вектор, а спин $\vec s$ -- псевдовектор (или аксиальный вектор). В
слабых взаимодействиях, обусловленных обменом $W$-бозонами,
участвуют левые электроны и правые позитроны, что демонстрирует
нарушение $C$-инвариантности: под действием преобразования $C$
левый электрон переходит в не взаимодействующий с $W$ левый
позитрон. Однако при последующем $P$-преобразовании левый позитрон
переходит в правый -- CP сохраняется. Ли и Янг высказали свою
гипотезу в 1956 году; в 1964 году в распадах (на этот раз)
нейтральных каонов было обнаружено нарушение CP.

$K^0$-мезон состоит из ($d\bar s$)-кварков, а $\bar K^0$ -- из
($s\bar d$)-кварков. Странность нарушается в слабом
взаимодействии, и во втором порядке по слабому взаимодействию
происходят переходы $K^0 \leftrightarrow \bar K^0$. При этом
состояниями с определенной массой являются их линейные
суперпозиции
\begin{equation}
K_1^0 = \frac{K^0 + \bar K^0}{\sqrt 2} \;\; \mbox{\rm и} \;\;
K_2^0 = \frac{K^0 - \bar K^0}{\sqrt 2} \;\; , \label{4.22}
\end{equation}
имеющие противоположные CP-четности: $CP(K_1^0) = +1$, $CP(K_2^0)
= -1$. В распадах на два пиона образующееся состояние имеет
$CP(2\pi) = +1$, поэтому такие распады в силу сохранения CP
возможны только для $K_1^0$, что делает его короткоживущим. Он
получил название $K_S$ от ``short life time''. $K_2^0$ распадается
по трехчастичным каналам, живет долго и называется $K_L$ от ``long
life time''. В 1964 году были найдены нарушающие CP распады $K_L
\to 2\pi$. Это редкие распады; так распадается один из примерно
500 $K_L$-мезонов. Причиной нарушения CP, приводящего к распадам
$K_L \to 2\pi$, является ненулевая фаза в амплитуде $K^0 - \bar
K^0$ перехода, меняющая знак при замене $K^0$ на $\bar K^0$.

В 2000-х годах было обнаружено нарушение CP в распадах
$B_d$-мезонов; скажем, $\Gamma(B_d \to \pi^+\pi^-)\neq \Gamma(\bar
B_d \to \pi^+ \pi^-)$. Здесь опять дело в фазе. Амплитуды распадов
имеют следующий вид:
\begin{eqnarray}
M(B_d \to\pi^+ \pi^-)& = & A + B e^{i(\delta + \varphi)} \;\; ,
\nonumber \\
M(\bar B_d \to\pi^+ \pi^-)& = & A + B e^{i(\delta - \varphi)} \;\;
. \label{4.23}
\end{eqnarray}
Отличие $\delta$ и $\varphi$ от нуля приводят к неравенству
$|M(B_d)|^2 \neq |M(\bar B_d)|^2$.

Следующий вопрос: где в лагранжиане Стандартной Модели содержится
эта нарушающая CP-симметрию фаза? Она содержится в той части
лагранжиана, которая дает массу кваркам, при учете наличия трех
фермионных поколений. Для одного поколения эти слагаемые были
обозначены $\tilde{\cal L}_3$, см. ф-лу (\ref{4.20}). В случае
нескольких поколений константы $f_e$, $f_d$ и $f_u$ заменяются
матрицами $N\times N$, где $N$ -- число поколений. Кварковая часть
имеет следующий вид:
\begin{equation}
{\cal L}_{3q} = f_d^{ik} \bar Q_i^\prime H d_{Rk}^\prime +
f_u^{ik} \bar Q_i^\prime \tilde H u_{Rk}^\prime \;\; .
\label{4.24}
\end{equation}
Значок ``$\prime$'' у кварковых полей подчеркивает, что эти
состояния не имеют определенной массы из-за недиагональных
массовых членов. Переход к массовым состояниям требует
диагонализации матриц $f_{u,d}^{ik}$. При этом одновременно
становится диагональной связь хиггсовского бозона с кварками.
Диагонализовать произвольную матрицу можно, домножая ее справа и
слева на две различные унитарные матрицы. Обратные матрицы,
действуя на кварковые поля, переводят их в состояния с
определенными массами:
\begin{equation}
d_R = D_R d^\prime_R \; , \;\; d_L = D_L d_L^\prime \; , \;\; u_R
= U_R u_R^\prime \; , \;\; u_L = U_L u_L^\prime \;\; ,
\label{4.25}
\end{equation}
где $D_R$, ..., $U_L$ -- унитарные $N \times N$ матрицы; в случае
трех поколений -- это матрицы 3$\times$3. Переписывая лагранжиан
Стандартной Модели в терминах полей $d_R$, ..., $u_L$ мы
обнаружим, что матрицы $D_R$, ..., $U_L$ сокращаются в силу своей
унитарности во всех членах лагранжиана кроме вершины испускания
$W$-бозона, которая приобретает следующий вид:
\begin{equation}
\frac{g}{\sqrt 2} \bar u_L U_L D_L^+ \gamma_\mu d_L W_\mu =
\frac{g}{\sqrt 2} \bar u_L K \gamma_\mu d_L W_\mu \;\; ,
\label{4.26}
\end{equation}
где унитарная матрица $K$ описывает смешивание кварков в
заряженном токе. Унитарная матрица $N \times N$ содержит $N^2$
независимых вещественных параметров: $N(N-1)/2$ углов (как и
ортогональная $N \times N$ матрица) и $N^2 - N(N-1)/2 = N(N+1)/2$
фаз. Часть фаз не является физическими (или наблюдаемыми); от них
можно избавиться, домножая кварковые поля $u_{Li}$ и $d_{Li}$ на
фазовые множители. Всего таких фаз 2$N$, однако домножение всех
нижних кварков на одну и ту же фазу преобразует матрицу $K$ так
же, как и домножение всех верхних кварков на противоположную фазу.
Итого, вращением кварковых полей можно убрать из матрицы $K$
$2N-1$ фаз, оставив в ней $N(N+1)/2 - (2N-1) = (N-1)(N-2)/2$
независимых фаз. Впервые одна фаза появляется в случае трех
поколений, наблюдаемых на опыте. Она и объясняет нарушение CP в
распадах $K$- и $B$-мезонов. Матрица $K$ называется матрицей
смешивания кварков Кабиббо--Кобаяши--Маскава. В случае двух
поколений фазы нет; CP-инвариантность не нарушена. Имеется один
угол смешивания $\theta_c$, введенный Кабиббо в 1963 году. Отличие
его от нуля делает возможным распады странных частиц, так как
заряженный слабый ток содержит слагаемое $\bar u_L \gamma_\mu$
($\cos\theta_c d_L + \sin \theta_c s_L$). Десятью годами позже, в
1973 году, Кобаяши и Маскава заметили, что в случае трех
кварк-лептонных поколений в матрице смешивания появляется CP
нарушающая фаза. Кварки и лептоны третьего поколения были открыты
после этой работы, и сейчас данные по нарушению CP в распадах $B$-
и $K$-мезонов подтверждают механизм Кобаяши--Маскава. Фаза и три
угла смешивания кварков сейчас измерены с хорошей точностью; мы
приведем их численные значения в следующей лекции.

Из обсуждавшихся дискретных симметрий не нарушено только
произведение CPT. Дело в том, что замену $\bar x, t \to -\bar x,
-t$ можно получить непрерывным вращением четырехмерного
пространства-времени, поэтому Лоренц-инвариантность обеспечивает
эту симметрию. Поворот стрелы времени превращает частицу в
античастицу, от положительной энергии мы переходим к
отрицательной, от волновой функции налетающей частицы -- к
волновой функции вылетающей античастицы. Поэтому преобразование PT
следует дополнить C, и мы приходим к симметрии CPT,
обеспечивающей, в частности, равенство масс и полных времен жизни
частиц и античастиц.

\newpage

\begin{center}

{\bf Лекция 5 \\
Нейтрино.}

\end{center}

\setcounter{equation}{0} \def\theequation{5.\arabic{equation}}

\bigskip

Нейтрино не участвуют в сильных взаимодействиях и не имеют
электрического заряда; они участвуют только в слабых
взаимодействиях. Рассмотрение начнем с реакции $pp \to D e^+
\nu_e$, являющейся основным источником солнечной энергии. Заметную
часть этой энергии уносят нейтрино. Учитывая, что $m_p = 938.3$
МэВ, $m_n = 939.6$ МэВ, $m_e = 0.5$ МэВ и энергия связи протона и
нейтрона в дейтроне $\varepsilon = 2.2$ МэВ, найдем, что энергия
нейтрино $E_\nu \la 0.4$ МэВ. Спектр нейтрино, рождаемых в ядерных
реакциях на Солнце, тянется до 14 МэВ. Для сечения рассеяния таких
нейтрино на нуклоне, используя $E_\nu \sim 1$ МэВ, получим:
\begin{equation}
\sigma_{\nu N} \sim G_F^2 E_\nu^2 \sim \frac{10^{-10}}{m_p^2}
\left(\frac{E_\nu}{m_p}\right)^2 \la 10^{-44} \;\; \mbox{\rm см}^2
\;\; . \label{5.1}
\end{equation}

Беря в качестве средней плотности Солнца 1 г/см$^3$, найдем, что
длина свободного пробега солнечных нейтрино по порядку величины
равна
\begin{equation}
l = 1/(n \sigma) \sim 1/(10^{24} \frac{1}{\mbox{\rm см}^3} \cdot
10^{-44} \; \mbox{\rm см}^2) \sim 10^{20} \; \mbox{\rm см} \;\; .
\label{5.2}
\end{equation}
Учитывая, что радиус Солнца $R_\odot \approx 7 \cdot 10^{10}$ см,
мы получим, что нейтрино пролетает, не рассеиваясь, $10^9$ Солнц!
Эта оценка показывает, насколько слабо взаимодействуют нейтрино и
как тяжело их детектировать.

Следующий вопрос -- есть ли у нейтрино масса. В теорию массу
нейтрино можно ввести так же, как мы делали это для кварков и
заряженных лептонов, добавив поля правых нейтрино. Однако до конца
80-х годов считалось, что нейтрино безмассовы, и поэтому правые
нейтрино не добавлялись в лагранжиан Стандартной Модели. Наиболее
точные верхние ограничения на массу электронного нейтрино
следовали из анализа спектра электронов, образуемых в
$\beta$-распаде трития. Тритий выделен малым энерговыделением,
равным 18 кэВ. Формула для корня квадратного из спектра
образующихся электронов носит название графика Кюри и имеет
следующий вид:
\begin{equation}
\sqrt{\frac{d N_e}{dE_e}} \sim \sqrt{(\Delta - E_e)\sqrt{(\Delta -
E_e)^2 - m_\nu^2}} \;\; . \label{5.3}
\end{equation}

Для выяснения вопроса о том, есть ли у нейтрино масса, исследуется
форма графика Кюри вблизи максимальной энергии электронов: она
линейна по $E_\nu$ в случае $m_\nu =0$ и имеет корневое поведение
для массивного нейтрино. Современное ограничение $m_{\nu_e} < 2$
эВ получено в экспериментах в Троицке и Майнце; будущий
эксперимент Катрин в Карлсруэ будет на порядок чувствительнее и
доведет верхнее ограничение до 0.2 эВ (либо обнаружит массу
нейтрино).

К настоящему времени во многих экспериментах обнаружены осцилляции
нейтрино, что доказывает наличие у них массы. Рассмотрим
осцилляции в случае смешивания двух сортов нейтрино, $\nu_1$ и
$\nu_2$:
\begin{eqnarray}
\nu_e & = & \cos\theta \nu_1 + \sin\theta \nu_2 \;\; , \nonumber
\\
\nu_\mu & = & -\sin\theta \nu_1 + \cos\theta \nu_2 \;\; ,
\label{5.4}
\end{eqnarray}
где $\theta$ -- угол смешивания нейтрино. Рождающееся в ядерной
реакции на Солнце $\nu_e$ следующим образом эволюционирует во
времени и пространстве:
\begin{eqnarray}
|\nu_e(t)> & = & \cos\theta |\nu_1(t)> + \sin\theta |\nu_2(t)> =
\nonumber \\
& = & \cos\theta e^{-i E_1 t + i p_1 x}|\nu_1 > + \sin\theta e^{-i
E_2 t + i p_2 x}|\nu_2 > \;\; , \label{5.5}
\end{eqnarray}
где за направление распространения нейтрино взята ось $x$. Точка
$x=0$ отвечает месту, где произошла ядерная реакция. Поэтому мы
должны положить $E_1 = E_2 \equiv E$, так как кинематически
запрещено рождение мюонов и $\nu_\mu$ в ядерных реакциях,
энерговыделение в которых никогда не достигает $m_\mu = 105$ МэВ
(мы рассматриваем осцилляции в приближении плоских волн;
рассмотрение с помощью волновых пакетов приводит к тем же формулам
для вероятности осцилляций). Перепишем (\ref{5.5}), используя
равенство $p_2-p_1 = -\frac{m_2^2 - m_1^2}{2E} \equiv -\Delta m^2
/(2E)$:
\begin{equation}
|\nu_e(t) = e^{-i E t + i p_1 x}\left[\cos\theta |\nu_1 >
+\sin\theta {\rm exp}(-i \frac{\Delta m^2}{2E}x) |\nu_2 >\right]
\;\; . \label{5.6}
\end{equation}

Вероятность того, что, пролетев расстояние $x$, электронное
нейтрино останется электронным нейтрино, дается следующей
формулой:
\begin{eqnarray}
P_{ee} & = & |<\nu_e | \nu_e(t)>|^2 = |\cos^2\theta +\sin^2\theta
{\rm exp}(-i \frac{\Delta m^2}{2E} x)|^2 = \nonumber \\
& = & 1-\sin^2 2\theta \sin^2\left(\frac{\Delta m^2}{4E} x\right)
\;\; . \label{5.7}
\end{eqnarray}
Если угол смешивания $\theta$ равен нулю или $\pi/2$, то
флэйворные состояния $\nu_e$ и $\nu_\mu$ имеют определенные массы,
и осцилляции не происходят.

Можно убедиться, что $P_{e\mu} = 1- P_{ee}$, где $P_{e\mu}$ --
вероятность обнаружить на расстоянии $x$ мюонное нейтрино.
Выражение для $P_{ee}$ удобно переписать в следующем виде:
\begin{equation}
P_{ee} = 1-\sin^2 2\theta \sin^2 1.27 \left[\frac{\Delta m^2
(\mbox{\rm эВ}^2)}{E(\mbox{\rm МэВ})} x(\mbox{\rm м})\right] \;\;
, \label{5.8}
\end{equation}
где энергия нейтрино измеряется в МэВ'ах, а расстояние -- в
метрах, что удобно для электронных нейтрино, образующихся в
ядерных реакциях. Для рождаемых на ускорителях мюонных нейтрино
энергию удобнее измерять в ГэВ'ах, а расстояние -- в километрах
(при этом коэффициент 1.27 в (\ref{5.8}) не изменяется).

Амплитуда осцилляций максимальна при $\theta = \pi/4$. Следует
иметь в виду, что в условиях реального эксперимента с реакторными
и ускорительными нейтрино или при наблюдениях солнечных и
атмосферных нейтрино происходит усреднение по некоторым интервалам
$x$ и $E$, что ведет к частичному замыванию осцилляционной
картины.

Уменьшение потока $\nu_e$ от Солнца объясняется осцилляциями;
осцилляции реакторных $\bar\nu_e$ были обнаружены в эксперименте
Kamland. Для угла смешивания и разности квадратов масс получено:
\begin{eqnarray}
\theta_{12} & \equiv & \theta_\odot = 35^o \pm 3^o \;\; ,
\nonumber \\
\Delta m_{21}^2 & \equiv & m_2^2 - m_1^2 = (8.0 \pm 0.3) \cdot
10^{-5} \mbox{\rm эВ}^2 \;\; . \label{5.9}
\end{eqnarray}

Согласно (\ref{5.8}) вероятность осцилляции не зависит от знака
$\Delta m^2$, однако влияние среды, в которой распространяются
нейтрино, позволяет этот знак определить (эффект
Михеева--Смирнова--Вольфенштейна). $\nu_2$ оказалось тяжелее
$\nu_1$. Недостаток солнечных нейтрино впервые был обнаружен в
70-х годах в эксперименте Дэвиса и потом был подтвержден в целом
ряде экспериментов. Исключительно интересные результаты в 2000-х
годах получены на детекторе SNO (Sudbury Neutrino Observatory,
Канада). Нейтрино рассеиваются в тяжелой воде, и обнаружен как
недостаток $\nu_e$ по реакции $\nu_e d \to e^- pp$, так и тот
факт, что $\nu_e$ осциллируют в активные в нейтральных токах
нейтрино: число реакций $\nu d \to \nu p n$, идущих за счет обмена
$Z$-бозонами, совпадает с ожидаемым в отсутствии осцилляций.

Существование нейтринных осцилляций считается твердо установленным
в 1998 году после публикации данных японского детектора
Суперкамиоканда, свидетельствовавших об осцилляции атмосферных
нейтрино. Эти нейтрино образуются при распадах $\pi$-мезонов и
мюонов, рождаемых при взаимодействии состоящих в основном из
протонов космических лучей с атмосферой Земли. В первичной реакции
происходит множественное рождение пионов. Основной распад
заряженных пионов $\pi^\pm \to \mu^\pm \stackrel{(-)}{\nu_\mu}$
сопровождается распадом мюона $\mu^\pm \to e^\pm
\stackrel{(-)}{\nu_e} \stackrel{(-)}{\nu_\mu}$.  Итого на каждое
электронное нейтрино (либо антинейтрино) приходится $\nu_\mu$ и
$\bar\nu_\mu$; отношение $[N(\nu_\mu) + N(\bar\nu_\mu)]/[N(\nu_e)+
N(\bar\nu_e)]$ должно быть близко к двум. Проведем мысленно
плоскость, перпендикулярную радиусу Земли, через детектор
Суперкамиоканда и разделим детектируемые нейтрино на летящих
``сверху'' и летящих ``снизу''. Тогда для первых соотношение 2:1
выполняется, в то время как ``снизу'' летит примерно одинаковое
количество $\nu_\mu$ и $\nu_e$. При этом поток $\nu_e$ ``сверху''
и ``снизу'' примерно одинаков и совпадает с теоретически
предсказанным; недостаток испытывает поток идущих ``снизу''
мюонных нейтрино. Поэтому объяснение следует искать в осцилляциях
$\nu_2 \leftrightarrow \nu_3$, так как в случае
$\nu_2\leftrightarrow \nu_1$ осцилляций поток $\nu_e$ увеличивался
бы. То, что осциллируют только идущие ``снизу'' $\nu_\mu$, говорит
о том, что длина осцилляций заметно больше толщи атмосферы $\sim$
10 км. Поэтому даже родившиеся в верхних слоях атмосферы $\nu_\mu$
не успевают проосциллировать до детектора. На ускорителях пучки
нейтрино создаются в распадах заряженных $\pi$ и $K$-мезонов, и
это пучки $\nu_\mu$ (или $\bar\nu_\mu$) с небольшой примесью
электронных нейтрино. Их осцилляции с такими же параметрами,
которые требуются для объяснения данных по атмосферным нейтрино,
наблюдались в Японии в эксперименте K2K (нейтрино рождались на
ускорителе KEK и детектировались детектором Суперкамиоканда,
расположенном на расстоянии примерно 250 км) и в США в
эксперименте MINOS (рождаемые в Батавии нейтрино детектировались
примерно в 700 км от ускорителя). В обоих ускорительных
экспериментах был обнаружен недостаток мюонных нейтрино. Все
обсуждавшиеся до сих пор эксперименты фиксировали частичное
``исчезновение'' $\nu_e$ или $\nu_\mu$ за счет осцилляций. В
эксперименте OPERA (рождаемые в ЦЕРН $\nu_\mu$ детектируются в
лаборатории Гран Сассо в Италии) ищут ``появление'' $\nu_\tau$. К
настоящему времени зафиксировано рождение в детекторе двух
$\tau$-лептонов за счет $\nu_\tau \to\tau$ перехода (заряженный
ток).

Из экспериментальных данных получены следующие параметры 2-3
осцилляций:
\begin{eqnarray}
\theta_{23} & = & 45^o \pm 4^o \;\; , \nonumber \\
|\Delta m_{23}^2| & = & (2.4 \pm 0.1) 10^{-3} \; \mbox{\rm эВ}^2
\;\; . \label{5.10}
\end{eqnarray}

Аналогичная матрице смешивания трех поколений кварков матрица
смешивания трех поколений лептонов получила название матрицы PMNS
в честь впервые рассмотревшего осцилляции нейтрино Понтекорво
(1957 г.), а также Маки, Накагавы и Сакаты. В случае дираковских
нейтрино она, так же как и матрица смешивания кварков,
параметризуется тремя углами и одной фазой и может быть получена в
результате перемножения трех матриц, описывающих вращения вокруг
третьей, второй и первой осей координат:
\begin{equation}
V_{PMNS} = \left(
\begin{array}{ccc}
1 & 0 & 0 \\
0 & c_{23} & s_{23} \\
0 & -s_{23} & c_{23}
\end{array}
\right)\left(
\begin{array}{ccc}
c_{13} & 0 & s_{13} e^{-i\delta} \\
0 & 1 & 0 \\
-s_{13} e^{i\delta} & 0 & c_{13}
\end{array}
\right) \left(
\begin{array}{ccc}
c_{12} & s_{12} & 0 \\
-s_{12} & c_{12}& 0 \\
0 & 0 & 1
\end{array}
\right) \;\; . \label{5.11}
\end{equation}

Угол $\theta_{13}$ был измерен одновременно в нескольких
экспериментах в 2012 году и оказался малым. Перемножая матрицы в
(\ref{5.11}), подставляя $\theta_{12} = \theta_\odot$,
$\theta_{23} = 45^o$ и предполагая малость $\sin\theta_{13}$,
получим:
\begin{equation}
\left(\begin{array}{c} \nu_e \\
\nu_\mu \\
\nu_\tau
\end{array}
\right) = \left(
\begin{array}{ccc}
c_\odot & s_\odot & s_{13} e^{i\delta} \\
-\frac{s_\odot}{\sqrt 2} & \frac{c_\odot}{\sqrt 2} &
\frac{1}{\sqrt 2} \\
\frac{s_\odot}{\sqrt 2} & -\frac{c_\odot}{\sqrt 2} &
\frac{1}{\sqrt 2}
\end{array}
\right) \left(
\begin{array}{c} \nu_1 \\
\nu_2 \\
\nu_3
\end{array}
\right) \;\; . \label{5.12}
\end{equation}
Здесь $\nu_1$, $\nu_2$ и $\nu_3$ -- состояния с определенными
массами, а $\nu_e$, $\nu_\mu$ и $\nu_\tau$ входят в заряженный
ток, будучи умноженными на $e$, $\mu$ и $\tau$, соответственно.
Иерархия $|\Delta m_{12}^2| \ll |\Delta m_{13}^2| \approx |\Delta
m_{23}^2|$ и $\theta_{13} \ll 1$ позволяла нам до сих пор
ограничиваться рассмотрением осцилляций в системе двух нейтрино.
Рождаемое в ядерных реакциях $\nu_e$ в пределе $s_{13} =0$ состоит
из двух массовых состояний. Учет отличия $s_{13}$ от нуля приводит
к ряби в вероятности $P_{ee}(x)$, длина волны которой в $|\Delta
m_{13}^2 /\Delta m_{12}^2 | \approx$ 30 раз меньше длины волны
осцилляций $\nu_e$, а амплитуда подавлена фактором $s_{13}^2$.
Именно эта ``рябь'' позволила измерить угол $\theta_{13}$ в
реакторных экспериментах Double Chooz, RENO и Daya Bay с
относительно коротким (порядка одного километра) расстоянием от
реакторов до дальних детекторов. Уменьшение потока $\bar\nu_e$ в
дальних детекторах по сравнению с потоком в ближних детекторах
демонстрирует осцилляции, отвечающие $|\Delta m_{13}^2| \simeq
|\Delta m_{23}^2|$, померенному ранее в осцилляциях атмосферных и
ускорительных $\nu_\mu$, и
\begin{equation}
\theta_{13} = 9^o \pm 1^o \;\; . \label{5.13}
\end{equation}

Этот результат был опубликован в 2012 году. Осцилляции
ускорительных и атмосферных $\nu_\mu$ происходят с относительно
малой длиной волны, и поэтому фаза между $\nu_1$ и $\nu_2$ на
таких расстояниях набежать не успевает, что оправдывает описание в
рамках двух массовых состояний нейтрино: $-s_\odot \nu_1 +
c_\odot\nu_2$ и $\nu_3$. Для флэйворного содержаний $\nu_3$ из
(\ref{5.12}) получим:
\begin{equation}
\nu_3 = s_{13} \nu_e + \frac{1}{\sqrt 2} (\nu_\mu + \nu_\tau) \;\;
, \label{5.14}
\end{equation}
поэтому в результате осцилляций с определяемой $|\Delta m_{23}^2|$
длиной волны ускорительных $\nu_\mu$ в них возникает примесь
$\nu_e$. Обнаружение этой примеси в 2011 году в эксперименте T2K
позволило заключить, что $\theta_{13}$ отличен от нуля. Измерение
его величины было опубликовано в 2013 году и дало близкий к
(\ref{5.13}) результат; задержка была связана с сильным
землетрясением 2011 года в Японии. Из параметров нейтринных
осцилляций осталось установить знак $\Delta m_{31}^2$: является ли
$\nu_3$ тяжелее $\nu_2$ и $\nu_1$ (так называемая нормальная
иерархия) или легче (обратная иерархия). Также остается определить
величину нарушающей CP фазы $\delta$, что можно сделать, сравнивая
вероятности $\nu_\mu \to \nu_e$ и $\bar\nu_\mu \to \bar\nu_e$
осцилляций. И, наконец, надо выяснить природу массы нейтрино и
измерить ее величину.

Нейтрино, будучи нейтральной частицей, может описываться
вещественным полем аналогично тому, как описывается фотон или
$\pi^0$-мезон. Такое нейтрино было рассмотрено еще в 30-х годах
Этторе Майорана и получило название майорановского нейтрино в
отличие от дираковского, описываемого комплексным полем. Матрица
смешивания майорановских нейтрино отличается от дираковских, так
как теперь мы теряем свободу домножать поле на фазу, и из матрицы
смешивания лептонов можно убрать только три фазы из шести
домножением на фазу полей заряженных лептонов. Из трех оставшихся
фаз одна называется дираковской, и это фаза $\delta$, уже
встречавшаяся нам. К ней добавляются две так называемые
майорановские фазы. В формулах для осцилляций начальные и конечные
состояния -- нейтрино с определенными флэйворами; массовые
состояния являются промежуточными. В результате майорановские фазы
входят в амплитуды в комбинациях вида $e^{i\alpha}\times
(e^{i\alpha})^*$ и выпадают. Они остаются в амплитудах $2\beta
0\nu$ распадов, обнаружение которых позволило бы доказать, что
нейтрино являются майорановскими частицами. Существуют ядра,
распадающиеся по схеме $N \to N^\prime 2e^- 2\bar\nu_e$,
называемой двойным $\beta$-распадом, или $2\beta 2\nu$-распадом. В
экспериментах детектируются электроны; их суммарная энергия
изменяется от $2m_e$ до полной энергии, равной разности масс
исходного и конечного ядер. Если нейтрино майорановские, то
лептонный заряд не сохраняется, и возможными становятся
безнейтринные $2\beta 0\nu$ распады: суммарная энергия электронов
равняется разности масс ядер, непрерывный спектр $dN/dE_e$
заменяется дельта-функцией. Амплитуда $2\beta 0\nu$ распада
пропорциональна линейной комбинации масс нейтрино, домноженных на
углы и фазы матрицы смешивания. Современное ограничение на эту
комбинацию близко к 1 эВ; планируемые эксперименты позволят
понизить это ограничение на порядок, либо обнаружить $2\beta 0\nu$
распад, что докажет майорановскую природу нейтрино.

В заключение этой лекции приведем численные значения параметров
матрицы смешивания кварков. Определяемый в слабых распадах
странных частиц $\sin\theta_{12} \approx 0.22$. В распадах
$B$-мезонов с образованием очарованных частиц измеряется угол
$\theta_{23}$: $\sin\theta_{23} \approx 4 \cdot 10^{-2}$. В
распадах $B$-мезонов без рождения чарма определяется
$\sin\theta_{13} \approx 4 \cdot 10^{-3}$. Наконец, вся
совокупность данных по нарушению CP с распадах $B$- и $K$-мезонов
описывается фазой $\delta \approx 70^o$. Видно, что матрица
смешивания кварков близка к диагональной, в отличие от матрицы
смешивания лептонов, в которой углы $\theta_{12}$ и $\theta_{23}$
велики.

\newpage

\begin{center}

{\bf Лекция 6 \\
За Стандартной Моделью.}

\end{center}

\setcounter{equation}{0} \def\theequation{6.\arabic{equation}}

\bigskip

Рассматривавшаяся нами до сих пор Стандартная Модель физики
элементарных частиц всесторонне экспериментально проверена и
надежно установлена. При любом дальнейшем развитии, при построении
новой, более совершенной, теории, Стандартная $SU(3)_C \times
SU(2)_L \times U(1)$ Модель никуда не пропадет и останется важной
частью новой теории. В этой лекции будут обсуждаться возможные
пути расширения Стандартной Модели, поэтому ее статус иной:
обсуждаются не твердо установленные факты, а некоторые спекуляции,
часть из которых, возможно, войдет в будущую теорию.

Естественно начать с очевидного недостатка Стандартной Модели: в
ней имеется слишком много свободных параметров, б\'{о}льшая часть
которых сосредоточена в юкавском секторе: 12 масс кварков и
лептонов, 8 (или 10 в случае майорановских нейтрино) углов и фаз
смешивания лептонов и кварков. Еще имеются три калибровочные
константы связи, масса и вакуумное среднее бозона Хиггса. Итого
около 25 параметров, что слишком много для фундаментальной теории.
Теории Великого Объединения связывают некоторые из этих
параметров; они основаны на объединении калибровочных
взаимодействий в одну группу: $SU(5)$, или $SO(10)$, имеются и
более экзотические возможности, основанные на исключительных
группах. Объединение констант в этих теориях происходит при очень
высоких энергиях $\sim 10^{15}$ ГэВ, что делает невозможной их
прямую экспериментальную проверку.

Наряду с обсуждавшимися взаимодействиями элементарных частиц, все
они участвуют в гравитационном взаимодействии. В силу лежащего в
основе общей теории относительности принципа эквивалентности
гравитирует не гравитационная масса, а тензор энергии-импульса
элементарных частиц, размерность которого равна четырем. Будучи
умноженным на поле гравитона $h_{\mu\nu}$ размеренности один, мы
приходим к размерной константе связи, равной $1/M_P$, где $M_P =
10^{19}$ ГэВ -- масса Планка. Такая теория неперенормируема: ряд
теории возмущений идет по $(E/M_P)^2$. При этом древесное
приближение работает прекрасно, правильно описывая отклонение
света в гравитационном поле Солнца и другие эффекты ОТО.
Непротиворечивое включение гравитации в физику элементарных частиц
является важнейшей задачей.

Третья проблема Стандартной Модели, которую мы собираемся
обсудить, носит называние проблемы иерархий, или проблемы тонкой
настройки. В лагранжиане Стандартной Модели есть один размерный
параметр: отрицательный квадрат массы бозона Хиггса, член $m_0^2
H^+ H$. Радиационные поправки к этому параметру расходятся и
содержат ряд по константе связи:
\begin{equation}
[m_0^2 + (c_1 g^2 + c_2 g^4 + ...)\Lambda^2]H^+ H \;\; .
\label{6.1}
\end{equation}
Стандартная процедура состоит в подборе затравочного параметра
$m_0^2$ таким образом, чтобы сумма в квадратных скобках
(\ref{6.1}) равнялась $m^2 \sim (100 \; \mbox{\rm ГэВ})^2$ --
величине, дающей наблюдаемое значение массы бозона Хиггса.
Проблема возникает, когда мы придаем параметру ультрафиолетового
обрезания $\Lambda$ физическое значение. Пусть он по порядку
величины равен масштабу Великого Объединения $10^{15}$ ГэВ, либо
массе Планка $10^{19}$ ГэВ. Тогда требуется чрезвычайно тонкая
настройка параметра $m_0$ с тем, чтобы при вычитании двух чисел,
каждое из которых $\sim 10^{30} \div 10^{40}$ ГэВ$^2$ осталось
$\sim 10^4$ ГэВ$^2$. Гораздо естественнее предположить, что
разность -- порядка самих этих чисел. Но тогда масштаб слабых
взаимодействий оказывается неприемлемо большим. В теория с
элементарными скалярными полями иерархия масштабов выглядит
чрезвычайно искусственно. Этой проблемы не возникает со спинорами
-- скажем, радиационные поправки к массе электрона зависят от
$\Lambda$ лишь логарифмически:
\begin{equation}
m_e = \left[1+(d_1 e^2 + d_2 e^4 + ...) \ln\frac{\Lambda^2}{m_{e_
0}^2}\right] m_{e_0} \;\; . \label{6.2}
\end{equation}
Даже при $\Lambda \sim 10^{19}$ ГэВ иерархии между $m_e$ и
$m_{e_0}$ не возникает; они имеют близкие численные значения.
Именно на этом наблюдении основано решение проблемы иерархий с
помощью низкоэнергетической суперсимметрии.

Суперсимметрией называется симметрия между бозонами и фермионами.
В супермультиплет входят бозоны и фермионы; в силу симметрии их
массы равны. Равны и радиационные поправки к массам; квадратичные
расходимости в массах скалярных частиц сокращаются, а остающиеся
логарифмические расходимости, как уже было сказано, не приводят к
проблеме иерархий.

Суперсимметричные теории поля исключительно красивы, и до сих пор
велика надежда на то, что низкоэнергетическая суперсимметрия имеет
место в природе. Отсутствие вырождения по массе известных
фундаментальных частиц с разными спинами говорит о том, что
суперсимметрия должна быть нарушена. В этих моделях
предсказывается большое количество новых частиц: каждая известная
частица имеет суперпартнера. Спинорные кварки и лептоны --
скалярные кваркино и лептино; векторные фотоны, глюоны и
$W$-бозоны -- спинорные фотино, глюино и $W$-бозино; скалярный
хиггс -- спинорное хиггсино. Поиск суперпартнеров уже долгие годы
является одной из основных задач для ускорителей, рассчитанных на
рекордные энергии. Работавший на полную энергию 8 ТэВ LHC поднял
нижнюю границу на массы кваркино и лептино до примерно 1.5 ТэВ. С
переходом на энергию 14 ТэВ доступно станет открытие кваркино и
глюино с массами вплоть до 3 ТэВ и, быть может,
низкоэнергетическая (так!) суперсимметрия будет наконец
обнаружена.

До сих пор мы говорили о внутренних проблемах Стандартной Модели и
способах их решения. На необходимость расширения Стандартной
Модели указывают также космологические данные. Лишь около 5\%
плотности энергии во Вселенной приходится на частицы Стандартной
Модели -- протоны и находящиеся, в основном, в ядрах гелия
нейтроны. В пять раз б\'{о}льшая плотность приходится на так
называемую темную материю -- нейтральные не участвующие в сильных
взаимодействиях частицы. К тому времени, когда температура ранней
Вселенной упала до примерно одного электронвольта и началась
рекомбинация (образование нейтрального водорода из электронов и
протонов), темная материя уже собралась в сгустки, послужившие
зародышами галактик. Нейтрино были тогда релятивистскими, сгустков
не образовывали, и поэтому они не годятся на роль темной материи.
Важно отметить, что в суперсимметричных расширениях Стандартной
Модели имеется подходящий кандидат на роль частиц темной материи.
Это легчайшая суперсимметричная частица, английская абревиатура
LSP. Как правило, она стабильна и нейтральна, являясь смесью
суперпартнеров бозона Хиггса, фотона и $Z$-бозона. Ее масса
порядка 100 ГэВ; вычисление современной концентрации LSP во
Вселенной дает необходимые 25\% общей плотности энергии при вполне
приемлемых параметрах моделей. Этот результат рассматривается как
сильный аргумент в пользу низкоэнергетической суперсимметрии.
Поиски частиц темной материи ведутся во многих лабораториях мира.

Остальные 70\% плотности энергии приходятся на космологическую
постоянную, так называемый $\Lambda$-член. Численно он равен
примерно $(10^{-3} \; \mbox{\rm эВ})^4$, и теоретически
воспроизвести это значение возможно только в теории, объединяющей
гравитацию и физику элементарных частиц. Дело в том, что без учета
гравитации $\Lambda$-член никак не проявляется физически, являясь
плотностью энергии вакуума. В электрослабой теории его
естественное значение $\sim\eta^4 \sim (100 \; \mbox{\rm ГэВ})^4$,
в КХД оно $\sim(\Lambda_{\mbox{\rm КХД}})^4 \sim (1 \; \mbox{\rm
ГэВ})^4$, в теории гравитации $\sim(M_P)^4 = (10^{19} \; \mbox{\rm
ГэВ})^4$. Задача на будущее -- понять, каким образом эти
гигантские числа компенсируются, и остается наблюдаемый
чрезвычайно маленький, но ненулевой $\Lambda$-член.

Генерация избытка барионов во Вселенной также требует расширения
Стандартной Модели. Заполняющее Вселенную реликтовое излучение
имеет планковский спектр с температурой $2.7^o$ K, что отвечает
плотности фотонов, примерно равной 400 фотонам на см$^3$.
Усредненная по Вселенной плотность протонов примерна равна 1
протону на м$^3$. Итого, на один протон сейчас приходится примерно
$10^9$ фотонов. Если идти вспять по времени, то это отношение
кардинально изменяется, когда температура Вселенной превышает
$\sim$ 1 ГэВ. При этом, скажем, в реакции $\gamma\gamma \to p\bar
p$ интенсивно рождаются протоны и антипротоны (при таких
температурах правильнее говорить о кварках). Соответствующие
реакции происходят очень быстро, и при $T\ga 1$ ГэВ
устанавливаются термодинамические распределения кварков и
антикварков. При этом их плотность близка к плотности фотонов
$\sim T^3$, отличаясь от нее только за счет того, что кварки --
фермионы. Мы должны предположить, что при этих высоких
температурах на $10^9$ фотонов приходится $10^9$ антикварков и
$10^9 +1$ кварк. Такое предположение выглядит чрезвычайно
искусственно; гораздо естественнее предположить, что Вселенная
родилась с нулевым барионным зарядом, а наблюдаемая асимметрия
(BAU, барионная асимметрия Вселенной) возникла динамически. Три
условия, необходимые для генерации BAU, были сформулированы А.Д.
Сахаровым в 1967 году, вскоре после открытия реликтового излучения
и CP-нарушения в распадах каонов. Во-первых, должен не сохраняться
барионный заряд -- иначе из начального состояния с $B=0$
невозможно получить состояние с $B\neq 0$. Во-вторых, должны
нарушаться C- и CP-симметрии -- иначе процессы с рождением
барионов и антибарионов будут иметь одинаковую вероятность, и
состояния с $B > 0$ получить не удастся. И, наконец, в системе
должно нарушаться термодинамическое равновесие: в равновесном
состоянии не важно, какая реакция идет быстрее -- все они успевают
произойти, и функции распределения кварков и антикварков
становятся Ферми--Дираковскими с нулевым химпотенциалом.

В Стандартной Модели реализуются все три условия Сахарова:
отклонение от равновесия достигается за счет расширения Вселенной
и/или фазового перехода с образованием конденсата Хиггса;
барионный заряд нарушается аномалией в дивергенции барионного
тока, обусловленной нарушением лево-правой симметрии в слабом
взаимодействии. Тем не менее, численно в Стандартной Модели BAU
оказывается исчезающе малой. Наиболее популярный в настоящее время
сценарий генерации BAU называется лептогенезисом. Стандартная
Модель расширяется введением очень тяжелых правых нейтрино, в
распадах которых генерируется лептонная асимметрия, которая
перерабатывается в барионную асимметрию процессами, идущими за
счет аномалии. Отметим, что введение тяжелых правых нейтрино также
позволяет объяснить малость масс наблюдаемых нейтрино,
оказывающихся по порядку величины равными $m_\nu \sim m_l^2/M_N$,
где $m_l$ -- массы заряженных лептонов ($e$, $\mu$, $\tau$), а
$M_N$ -- массы правых нейтрино. Соответствующий механизм получил
название механизма качелей (see-saw).

Конечно нельзя исключить, что выход за рамки Стандартной Модели
будет осуществлен по сценарию, отличному от обсуждавшихся в этой
лекции. Быть может, по такому, о котором сегодня никто и не
подозревает.

Я благодарен организаторам Школы за приглашение прочитать лекции и
Е.А. Ильиной за помощь в оформлении рукописи.

\end{document}